\documentclass[article]{aastex63}
\usepackage{amsmath}
\usepackage{multirow}
\usepackage[utf8x]{inputenc}
\usepackage{float}
\usepackage{enumerate}
\usepackage{graphicx}
\usepackage{url}
\usepackage{hyperref}

\newcommand{\gsim}{\raisebox{-0.13cm}{~\shortstack{$>$ \\[-0.07cm]
      $\sim$}}~}
\usepackage{tabularx}
\usepackage{array}
\submitjournal{AJ}

\shorttitle{Collision Probabilities in the Kuiper belt.}
\shortauthors{Abedin et al.}

\begin{document}

\title{Collision Probabilities in the Edgeworth-Kuiper belt}

\correspondingauthor{Abedin Abedin}
\email{Abedin.Abedin@nrc-cnrc.gc.ca}

\author[0000-0002-7025-0975]{Abedin Y. Abedin}
\affiliation{National Research Council of Canada, Herzberg Astronomy and Astrophysics, 5071 West Saanich Road, Victoria, BC V9E 2E7, Canada}

\author[0000-0001-7032-5255]{JJ Kavelaars}
\affiliation{National Research Council of Canada, Herzberg Astronomy and Astrophysics, 5071 West Saanich Road, Victoria, BC V9E 2E7, Canada}

\author[0000-0002-4439-1539]{Sarah Greenstreet}
\affiliation{B612 Asteroid Institute,
	20 Sunnyside Avenue, Suite 427, Mill Valley, CA 94941, USA}
\affiliation{DIRAC Center, Department of Astronomy, University of Washington, 3910 15th Avenue NE, Seattle, WA 98195, USA}

\author[0000-0003-0407-2266]{Jean-Marc Petit}
\affiliation{Institut UTINAM, CNRS-UMR 6213, Université Bourgogne Franche Comté BP 1615, 25010 Besançon Cedex, France} 

\author[0000-0002-0283-2260]{Brett Gladman}
\affiliation{Department of Physics and Astronomy, 6224 Agricultural Road, University of British Columbia, Vancouver, British Columbia, Canada}

\author[0000-0001-5368-386X]{Samantha Lawler}
\affiliation{Campion College and the Department of Physics, University of Regina, Regina SK, S4S 0A2, Canada}

\author[0000-0003-3257-4490]{Michele Bannister}
\affiliation{University of Canterbury, Christchurch, New Zealand}

\author{Mike Alexandersen}
\affiliation{Minor Planet Center, Smithsonian Astrophysical Observatory, 60 Garden Street, Cambridge, MA 02138, US}

\author{Ying-Tung Chen}
\affiliation{Academia Sinica Institute of Astronomy and Astrophysics, Taiwan}

\author{Stephen Gwyn}
\affiliation{National Research Council of Canada, Herzberg Astronomy and Astrophysics, 5071 West Saanich Road, Victoria, BC V9E 2E7, Canada} 

\author[0000-0001-8736-236X]{Kathryn Volk}
\affiliation{Lunar and Planetary Laboratory, University of Arizona: Tucson, AZ, USA}

\begin{abstract}
Here, we present results on the intrinsic collision probabilities, $ P_I$, and range of collision speeds, $V_I$, as a function of the heliocentric distance, $r$, in the trans-Neptunian region. 
The collision speed is one of the parameters, that serves as a proxy to a collisional outcome e.g., complete disruption and scattering of fragments, or formation of crater, where both processes are directly related to the impact energy. We utilize an improved and de-biased model of the trans-Neptunian object (TNO) region from the ``Outer Solar System Origins Survey" (OSSOS). It provides a well-defined orbital distribution model of TNOs, based on multiple opposition observations of more than 1000 bodies. In this work we compute collisional probabilities for the OSSOS models of the main classical, resonant, detached+outer and scattering TNO populations. The intrinsic collision probabilities and collision speeds are computed using the \"{O}pik's approach, as revised and modified by Wetherill for non-circular and inclined orbits. The calculations are carried out for each of the dynamical TNO groups, allowing for inter-population collisions as well as collisions within each TNO population, resulting in 28 combinations in total. Our results indicate that collisions in the trans-Neptunian region are possible over a wide range in ($r, V_I$) phase space. Although collisions are calculated to happen within $r\sim 20 - 200$~AU and $V_I \sim 0.1$~km/s to as high as $V_I\sim9$~km/s, most of the collisions are likely to happen at low relative velocities $V_I<1$~km/s and are dominated by the main classical belt.
\end{abstract}



\section{Introduction}\label{sec:intro}
The Edgeworth-Kuiper belt 
(Kuiper belt hereafter) is a torus-shaped agglomeration of small icy objects, encircling the Solar System. 
Extending from the orbit of Neptune at 30~AU to more than 200~AU from the Sun, the Kuiper belt consists of more than $\sim10^5$ trans-Neptunian objects (TNOs) with size $D>100$~km \citep{petit11}. 
TNOs can be considered to consist of pristine building blocks, leftover from Solar System formation; they are unlikely to have been exposed to temperatures of $T>50$~K, \citep[e.g.][]{jewitt04} over the age of the solar system. 

Collisions between bodies inevitably lead to ejecta, regardless of whether the impact is disruptive or crater forming, with the total mass and velocity distribution of liberated material being dictated by the energy of the impact \citep[e.g.][]{melosh89,fuji89,hous90,benz99,hols02}. 
The ejecta size distribution generally follows some power-law, ranging from large boulders to dust particles, with the number of released fragments being inversely proportional to particle size. 
Understanding the impact velocity distribution and probability will aid in understanding the impact environment of these bodies and the expected rate of dust/debris production. 
These impacts may very well be responsible for formation of dust rings and satellites around planets \citep{harris84,charnoz09,charnoz18}, dusty exospheres around small bodies \citep[e.g.,][]{szalay18}, serve as source and replenishment of the interplanetary dust flux \citep[e.g.][]{kuchner10,vitense12,poppe16} or lead to planetary surface cratering \citep[e.g.][]{green15,green19}.  In particular, dust is dynamically short lived and collisions provide an avenue to refresh that content. 


Owing to the fundamental importance of the collisional processing for understanding the Kuiper belt and the outer solar system, a number of very early estimates of collisional environment were made with only the slimmest of observational data to ground them.
\citet{stern95} argued that TNOs smaller than $\sim$ 10 km are collisionally evolved, based on collision rate estimation between 35-70 AU, while objects with radii $r > 100$km may retain their primordial sizes. 
However, the only available observational statistics at the time was 25 objects of size similar to 1992 QB$_1$ and the rate of observed Jupiter-Family Comets (JFCs), assuming they originate from the low inclination component of the present Kuiper belt. 
\citet{davis97} argued that KBOs with diameters  ($D < 20$~km), are significantly collisionally evolved, producing on average 10 objects per year of size 1-10 kilometers. 
\citet{durda00} expanded on previous works using an updated number of discovered KBOs and estimated that 2.5 km KBOs have a collisional life-time of 3.5 Gyr, while sub-kilometer objects collide with 100 km TNOs once every $10^7 - 10^8$ years.
In the intervening decades since these early works a much improved picture of the orbital structure and size distribution of the Kuiper belt has emerged.

Surveys for Kuiper belt objects \citep[e.g.][]{jewitt93,jewitt95,gladman98,chiang99,glad01,trujillo01,millis02,jones06,jj09,Schwamb10,petit11,brown15,alexandersen16,banister16,banister18} have demonstrated that the Kuiper belt contains a rich diversity of orbital sub-components. Contemporary models of the formation of the Kuiper belt imply that Neptune has played a significant role in placing and reshaping the orbits of these primordial bodies \citep{malhotra93,malhotra95,levison03,gomes05,tsiganis05,nes15a,shannon19} with the orbital evolution of Neptune ``fossilized" into the current state of the Kuiper belt. 
Furthermore, studies of size-frequency distributions (SFDs) of the TNO sub-populations \citep[e.g.][]{jewitt95,gladman98,glad01,trujillo01,bernstein04,fras08a,fras08b,petit11,fras14,lawler18} provide clues to the agglomeration and collision history.  Studied as a whole, the Kuiper belt provides many insights into the processes of planet formation and planetary system evolution. 

With the increased number of discovered TNOs more complete Kuiper belt dynamical and physical models have become possible allowing more precise determination of collision probabilities. 
Using the CFEPS L7 model \citep{petit11}, \citet{delloro13} calculated the collision probabilities between different TNO populations and sub-populations, finding intrinsic collision probabilities to be lower than previous estimates.  Using refined versions of the L7 model, and informed by various luminosity function studies, \citet{green15,green16} made predictions for the collision probabilities and cratering rate onto Pluto and Charon prior to the NASA New Horizons Pluto flyby and later made predictions for New Horizons Kuiper belt Extended Mission target ``Arrokoth" (2014~MU$_{69}$) \citep{green19}. 



There are over 1200 KBOs in the OSSOS ensamble sample \citep{banister18}, with high-precision orbital elements and detailed discovery characterization.  This sample allows for the construction of a `de-biased' TNO model, using the OSSOS survey simulator \citep{petit11,Lawler18b}.  This improved knowledge of the structure and history the Kuiper belt enables a yet more precise determination of the collisional environment. 

We present here our determination of the intrinsic collision probabilities between the orbital sub-populations of the Kuiper belt based on this debiased model.  
The intrinsic probability is a function of the orbital distribution of objects.   
The orbital distribution of different dynamical TNO classes is the foundation for computing how frequently these objects collide.  When combined with a TNO size frequency distribution model the intrinsic probabilities determine a collision rate.
The goal of this work is to use the most up-to-date model of the Kuiper belt orbit distribution to obtain refined collision probabilities and speeds. 

In Section~\ref{sec:struc} we describe the orbital and size distributions used in our probability calculations.  Section~\ref{sec:colprob} presents our approach in determining the intrinsic and total collision probabilities in the current Kuiper belt.  In Section~\ref{sec:results} we present the intrinsic collision probabilities, considering each KBO sub-population in-turn and then combining those into a total collisional probability for objects in the Kuiper belt. In Section~\ref{sec:discus} we discuss some zeroth-order implications determined from  our collisional probabilities. 


\section{Dynamical Structure of the Kuiper belt} \label{sec:struc}

Determination of orbital elements for TNOs depends on accurate and multi-opposition astrometry, which is often challenging due to the large distances and faintness of TNOs \citep[e.g.][]{jones06}. 
The apparent daily motion of a TNO, at a distance of $r \sim 40$~au, is small (mean motion $n \sim 6.8\times10^{-5}$~au/day) and hence the observed arc ($\theta \sim 0.2$~arcmin), even over a month, will be a tiny fraction of the orbit. 
Often, observations over the course of a number of years are needed before the precise orbit of the TNO can be determined \citep{banister18}.
Another difficulty is introduced from the $r^{-4}$ distance dependence of the reflected light from a KBO, registered by telescopes. Contemporary KBO detection is limited to objects with $R-$band magnitude brighter than $m_R \lesssim 26.5 $ \citep{petit06,fras08b} (diameter roughly $D > 30$~km), although there have been reports of detection of smaller $D\sim 1$~km KBOs using stellar occultation \citep{schlicht09,aramitsu19}. 
Furthermore, the rate of detection per unit area is a strong function of pointing direction: TNOs moving on highly inclined orbits have a better chance of being discovered away from the ecliptic while those on low-inclination orbits are only detectable when the observed field is near the plane of the TNO's orbit. 
Similarly, TNOs on eccentric orbits are more difficult to detect, brightening sufficiently only near their perihelia, but spending most of their time near aphelia. 
Conversely, objects on more circular, moderate inclination orbits are more readily detected. 
Therefore a dedicated TNO survey is more suitable and desired for extensive follow-up observations, as well as characterization of the above biases. 
For a review on the biases associated with KBO observations, see \citet{kavelaars20}.

Nevertheless, since the discovery of the first KBO \citep{jewitt93}, there have been a number of KBO survey programs, among them \citet{trujillo01}, \citet{Schwamb10}, Deep Ecliptic Survey \citep{millis02},  Canada-France Eclipitc Plane Survey (CFEPS) \citep{jones06,jj09,petit11,gladman12}, \citet{alexandersen16} (MA) and the Outer Solar System Origin Survey (OSSOS) \citep{banister16,banister18,lawler18}. 
Presently, there are $\gsim 3300$ cataloged TNOs in the Minor Planet Center (MPC) \url{https://minorplanetcenter.net/iau/mpc.html}, but only half of them have sufficiently well-constrained orbits to allow for precise long term numerical integration of their orbits and accurate dynamical classification, with the majority of those coming from the OSSOS Ensemble sample which combines the CFEPS, MA and OSSOS samples. 
Therefore, we choose to utilize the most up-to-date bias-free TNO dynamical model by OSSOS as constrained by the OSSOS Ensemble of over 1000 TNO orbits \citep{banister18}. 

For this work we consider the collisional circumstances for different orbit classes of TNOs.  
We group the TNOs into dynamical groups proposed by \citet{glad08}, based on long-term numerical orbital integration.
\begin{enumerate}
	\itemsep0em
	\item Resonant objects - TNOs that are currently in mean motion resonance (MMR) with Neptune. 
	\item Scattering objects - TNOs that experience close encounters and ``scattering" off Neptune over 10 Myr of numerical orbital integration, 
	\item Classical and Detached objects - TNOs which do not fall into either of the above categories and is often further sub-divided on semi-major axis and inclination.
        \begin{enumerate}
        	\itemsep0em
        	\item Inner belt - classical objects which are interior to the 3:2 MMR with Neptune.
        	\item Main belt - The main classical belt KBOs, which have semi-major axes between the 3:2 and 2:1 MMR with Neptune. 
        	The main belt contains at least two distinct sub-classes: ``Hot" and the ``Cold" components, where the Cold appears to be constrained to a narrow ring between 42.5~au and 44.5~au and exhibits complex substructure. 
        	The Hot component is distinguished by a wide inclination ($i$) distribution whereas the Cold component's inclinations are tightly confined.
        	\item Outer belt - Objects with semi-major axes beyond the 2:1 MMR and eccentricities $e<0.24$.
        	\item Detached belt -  Objects with semi-major axes beyond the 2:1 MMR and eccentricities $e>0.24$.
        \end{enumerate}
\end{enumerate}
To compute the collision probabilities in the Kuiper belt we require models that describe the orbital distributions of each of these Kuiper belt classes.  

\subsection{Resonant class}
As we are interested in the global collision rates we only consider those resonances with a significant number of known members and a reliable orbit distribution model available in the literature.
For the sake of simplifying the computational problem, without significant loss in collisional probability precision we compute the probabilities grouped by semi-major axis:
\begin{enumerate}
    \item Inner resonances, in order of increasing semi-major axis, N4:3, N3:2
    \item Main resonances, (semi-major axis within the main belt resonances), N5:3 and N7:4
    \item Outer resonances N2:1, N7:3 and N5:2
\end{enumerate}
Where 'N' stands for a MMR with Neptune and the convention used is $p:q$, where $p$ and $q$ are the small integers that refer to the number of orbits completed by the inner body (Neptune in our case) and the outer body (TNO) respectively, where $p-q$ is referred to as the ``order of resonance". 
Our collisional modelling does not account for the full resonant dynamics of Kozai-type orbital oscillation nor the confined precision of the argument of pericenter as we treat these objects as having uniformly precessing angular elements. However, \citet{green15} find that accounting for the resonant and Kozai dynamics of the plutinos lead to an increased collision probabilities and rates onto Pluto by $\sim$50\%, though as will be shown later, neglecting the non-uniform precession of the orbital elements of resonant TNOs will not change our overall conclusion. 

\subsection{Scattering}
The orbital distribution of the scattering component can be determined via comparison with cosmogonic models of the formation of the Kuiper belt and Oort Cloud.  By definition, the orbital elements of the Scattering objects are `actively` evolving.  \citet{shankman16} and \citet{lawler18} found that the currently observed orbital distributions of the scattering members of the OSSOS ensemble sample are well represented by the output of simulations in \citet{kaib11}.  We utilize the orbital distributions provided by \citet{lawler18} as our orbit model for the Scattering component.

\subsection{Classical Kuiper belt}
Here, rather than consider parametric distributions of sub-components of the Classical Kuiper belt we utilize an orbit model derived by debiasing a detected sample. 
In this way the detected Classical belt objects and the discovery survey's efficiency of detecting such orbits are used to invert detected population of KBOs to determine the intrinsic population.  For this work we utilized the debiased model described in \citep{petit20}. 

Following \citep{petit20} we split the main classical belt on the basis of orbital inclination; Cold-KBOs with $i< 5^{\circ}$ and Hot-KBOs with $i\geq 5^{\circ}$. 
Here, ``Cold" and ``Hot" refer to dynamical state of the orbits, and are not related to the objects' temperature. 
This somewhat crude splitting is motivated by the fact that, although there is an overlap between Cold and Hot objects in the eccentricity and inclination space, one can see a clear concentration of objects with low ($i<5^{\circ}$) inclination \citep[e.g.][]{brown01,petit11}. 
This inclination cutoff is combined with low ($e<0.08$) eccentricity threshold to select out those low-i KBOs that are more strongly interacting with Neptune and group them with the Hot objects for the purpose of considering average collision probabilities.
We thus have two distinct TNO populations in the main classical belt, which allows us to test which of these (``Hot" vs. ``Cold") are more likely to experience collisions in the Kuiper belt. 

\subsection{Outer/Detached belt}

Significantly fewer non-resonant members of the Kuiper belt with $a > 47~au$, compared to $a < 47$~au, have been detected, making a direct debiasing statistically less certain.  For this region we revert to the CFEPS L7 \citep{petit11} parametric model.  
The L7 model represents the orbit distribution of this component as having $e$ and $i$ that follow the Hot Main Kuiper belt population but with $a$ following a probability density function (PDF) $\propto a^{-\frac{5}{2}}$ and the remaining elements drawn randomly.   
The $a$ PDF is normalized such that the Outer/Detached component represents a smooth continuation from the Hot Main Kuiper belt.  
As we will see, collisions among this population are rare and the precise orbital distribution is not likely to significantly modify conclusions about impact rates.

The seven main populations considered here - Cold and Hot Classicals,  Inner Resonant, Main Resonant, Outer Resonant, Detached+Outer belt and Scattering disk we have a total of 28 orbit class combinations. 

\section{Collision Probabilities}\label{sec:colprob}

We compute the intrinsic collision probability $P_I$ and speeds $V_I$ between each pair of orbits within, as well as between different dynamical classes.
In order to compute the intrinsic collision probability between each and every TNO dynamical class, we randomly select a sample of $10^4$ orbits from orbital distributions in each of the sub-population.
%
The intrinsic collision probability is independent of the sample size but larger samples provide somewhat better sampling of the available phase space for the orbit class being considered.  
For each orbit family pair we computed the intrinsic collision probability between each pair of orbits in each population:  the number of calculations roughly scales as $\sim N^2$, where $N$ is the number of individual orbits considered.
Each probability computation required an average CPU time of $t\sim 1.5$~ms per orbit pair.
Performing tests with different sample sizes, we concluded that computation between $10^4$ orbits, which is equivalent to $\sim 5\times10^7$ orbit pairs (requiring about 20 hours of computing) provided statistically equivalent behaviour to a $10^5$ sample (requiring about 200 hours of computing). We then compute the intrinsic collision probabilities between each of the 28 permutation of our orbit classes.   


We use the approach described in \citet{weth67} to compute the intrinsic collision probability between two orbits:
\begin{equation}\label{eq:intr}
\begin{aligned}
		 P_I &= \frac{V_I R_h}{8\pi^2 \sin i' |\cot\alpha| a_1a_2\sqrt{p_1p_2}},\\
		 p_1 &= a_1(1-e_1^2)\\
		 p_2 &=a_2(1-e_2^2)\\
		 R_h &= \frac{p_2}{1+e_2\cos\omega}\\
		\cot\alpha &= R_h\dot{\omega}/\dot{R_h}  
\end{aligned}
\end{equation}
\noindent where $p_1$ and $p_2$ are the semi-latus rectums of the two orbits, $i'$ is the mutual inclination and $V_I$ is the relative velocity between the two bodies at the point of orbit intersection. 
The distance $R_h$ is the heliocentric distance of the intersection point (mutual node) of the two orbits and $\omega$ is the orientation of perihelion of the impacting orbit and has a meaning of true anomaly of the intersection point. 
$\alpha$ is the angle between object's rectilinear trajectory and the heliocentric distance vector. 
The intrinsic collision probability is a function of the orbital elements of the given orbit pair only, without involving their size.
Due to gravitational perturbations, the arguments of pericenter and the longitudes of the ascending node of the two orbits will experience secular variations, causing the mutual inclination between the two orbits to process.  
Eq.~\ref{eq:intr} does not incorporate the longitude of the ascending node and the argument of pericenter, as the formalism of \citep{opik51} and \citep{weth67} assumes that all values for these parameters, during one procession cycle, are equally probable\footnote{We do not consider the effects of resonant dynamics of the colliding objects as these do not strongly alter inter-class collisional probabilities.}. 

For the computations we make use of the FORTRAN implementation used by \citet{green15} to calculate the collision probabilities and speeds between each orbit pair within and between different TNO populations.  
However, we extended this code to provide the intrinsic collision probability as both a function of speed (as in \citet{green15}) and also the heliocentric distance at time of collision. 
We provide our results as a matrix of 200$\times$135 distance, velocity cells. 
For each cell we compute the average collision probability within a 1~au, 1~m/s volume at a the heliocentric distance $R$ with an impact speed $V_I$ for that cell. 
The full grid of collision probabilities is provided in electronic form.



\begin{figure*}[htpb]
	\gridline{\fig{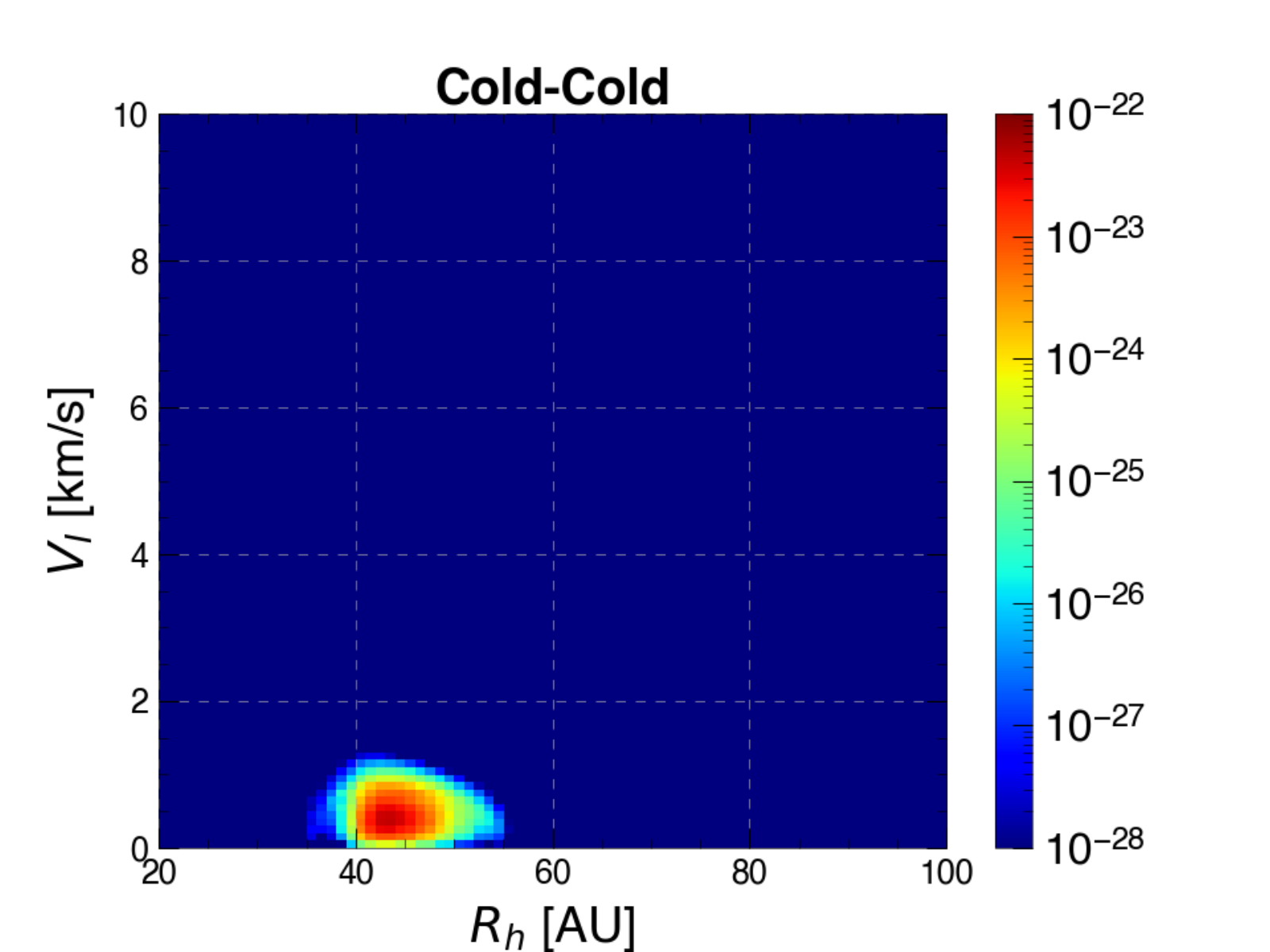}{0.33\textwidth}{(a)}
		\fig{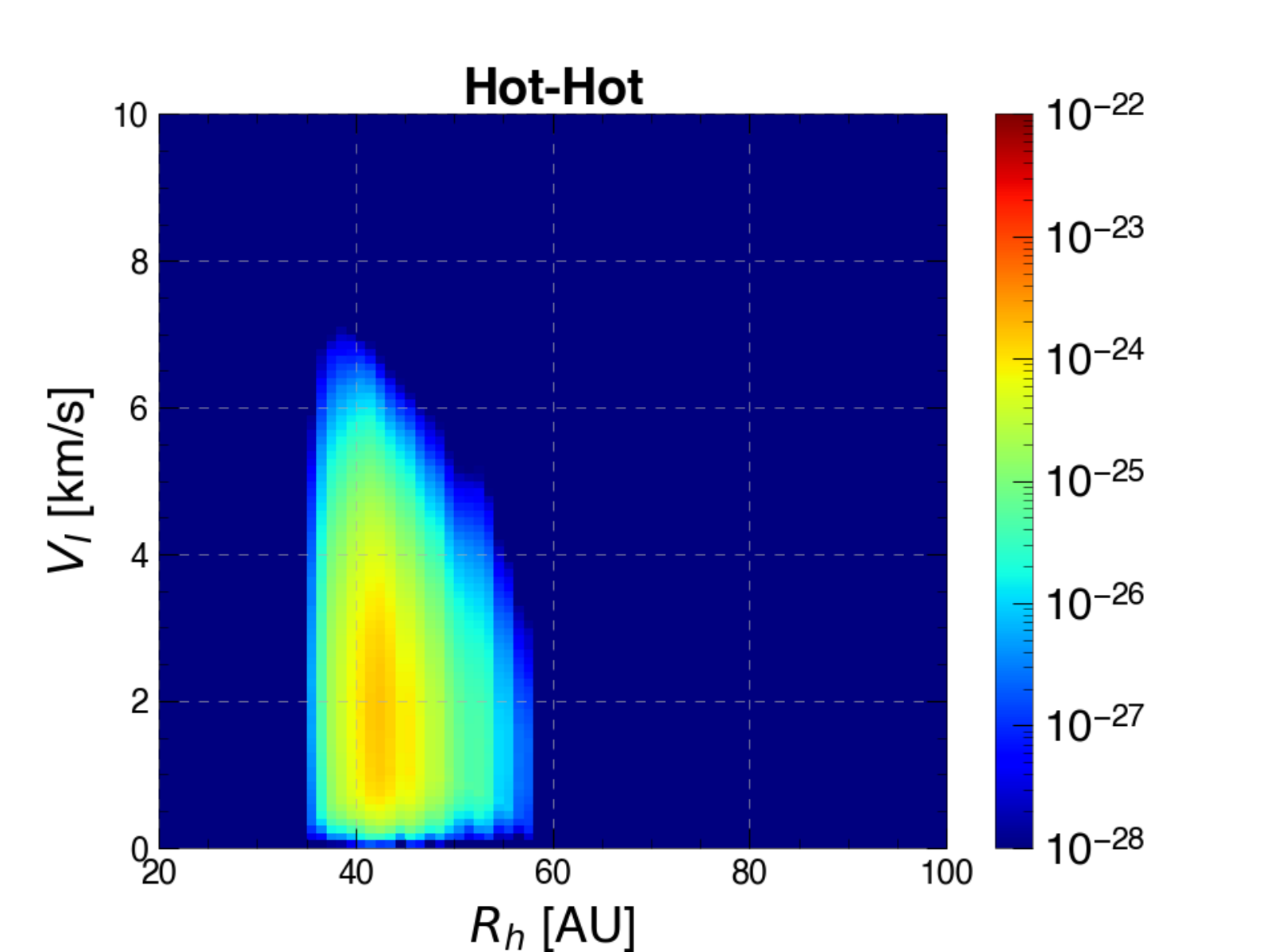}{0.33\textwidth}{(b)}
		\fig{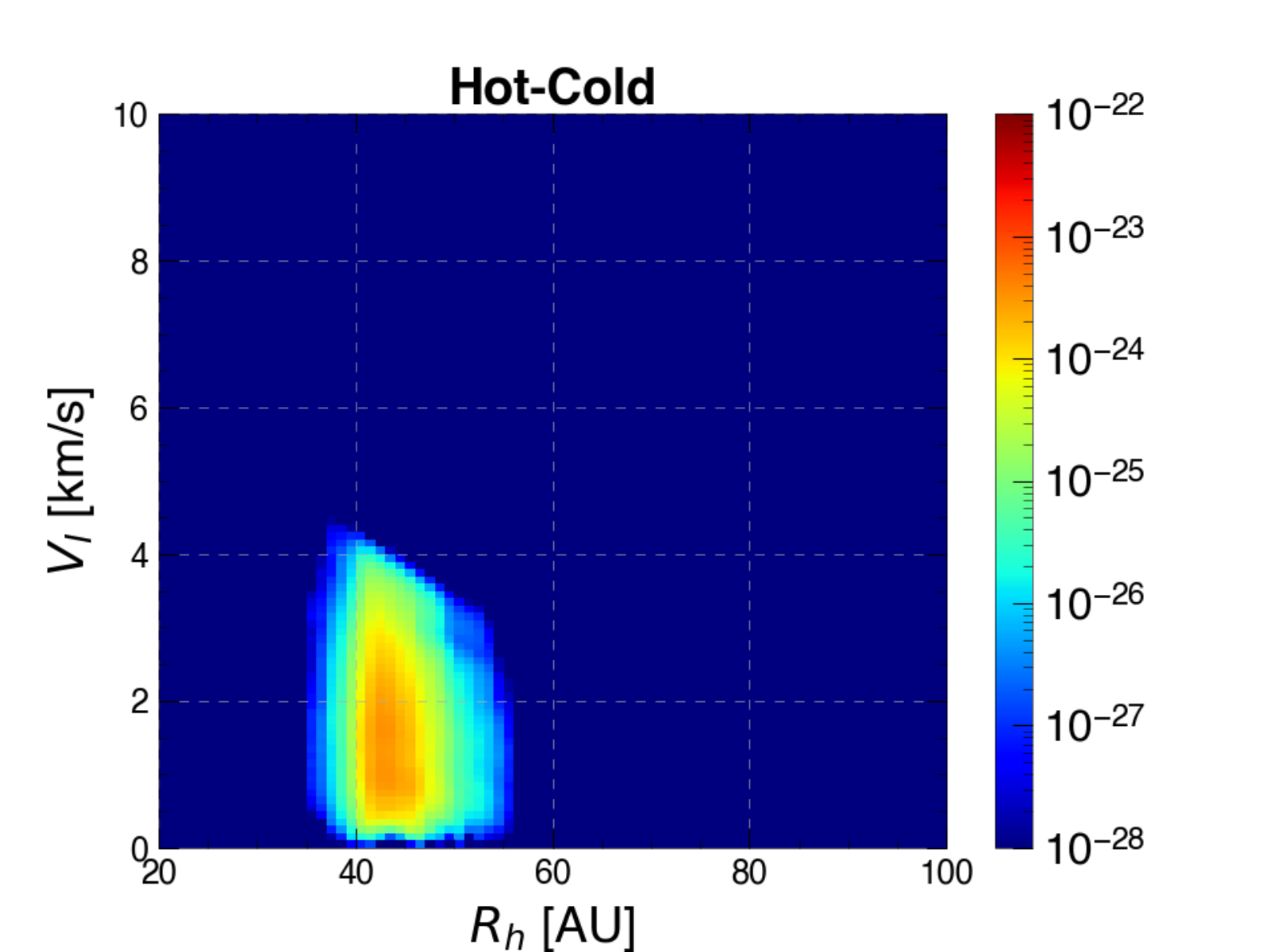}{0.33\textwidth}{(c)}}
	\gridline{\fig{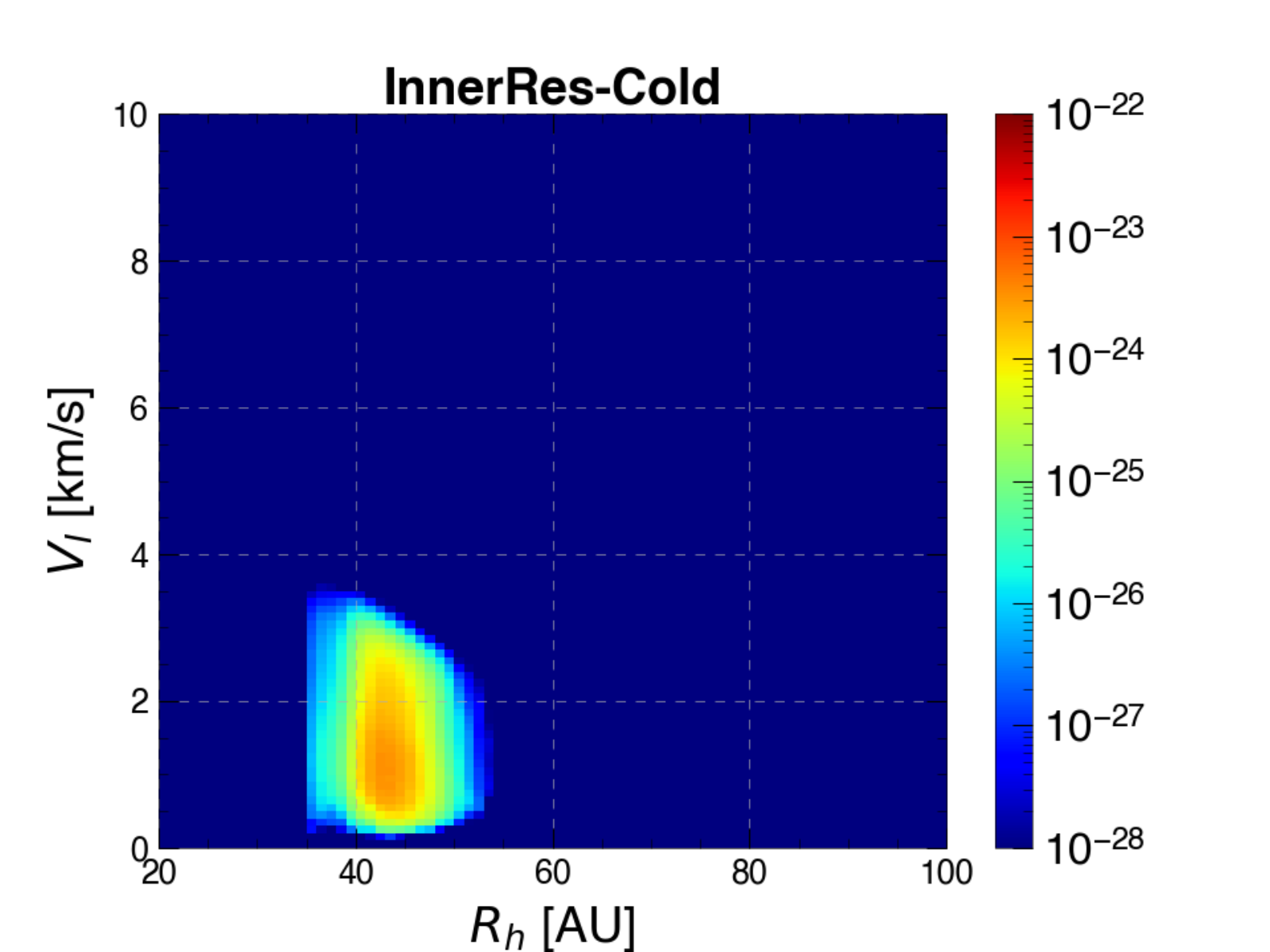}{0.33\textwidth}{(d)}
		\fig{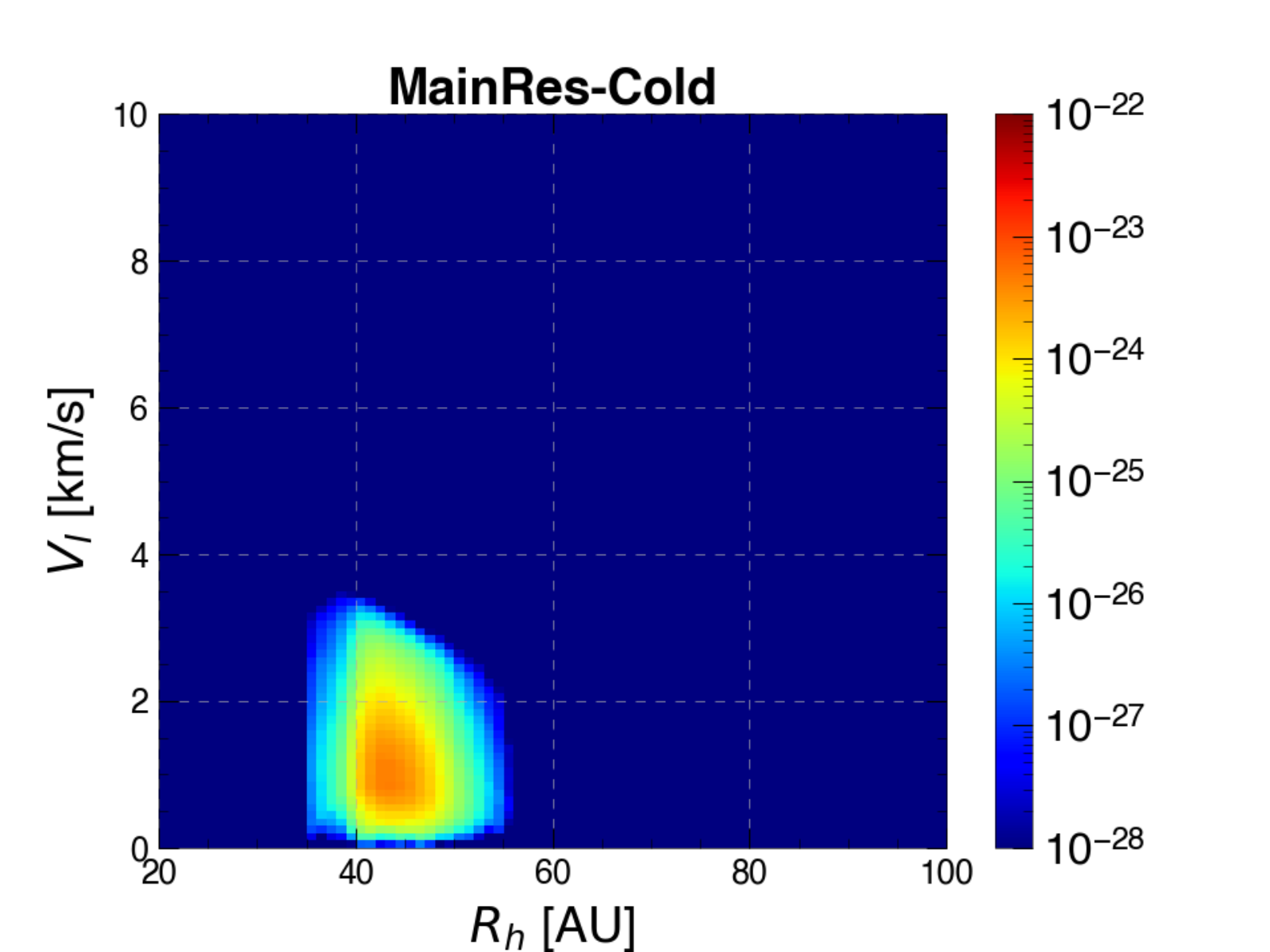}{0.33\textwidth}{(e)}
		\fig{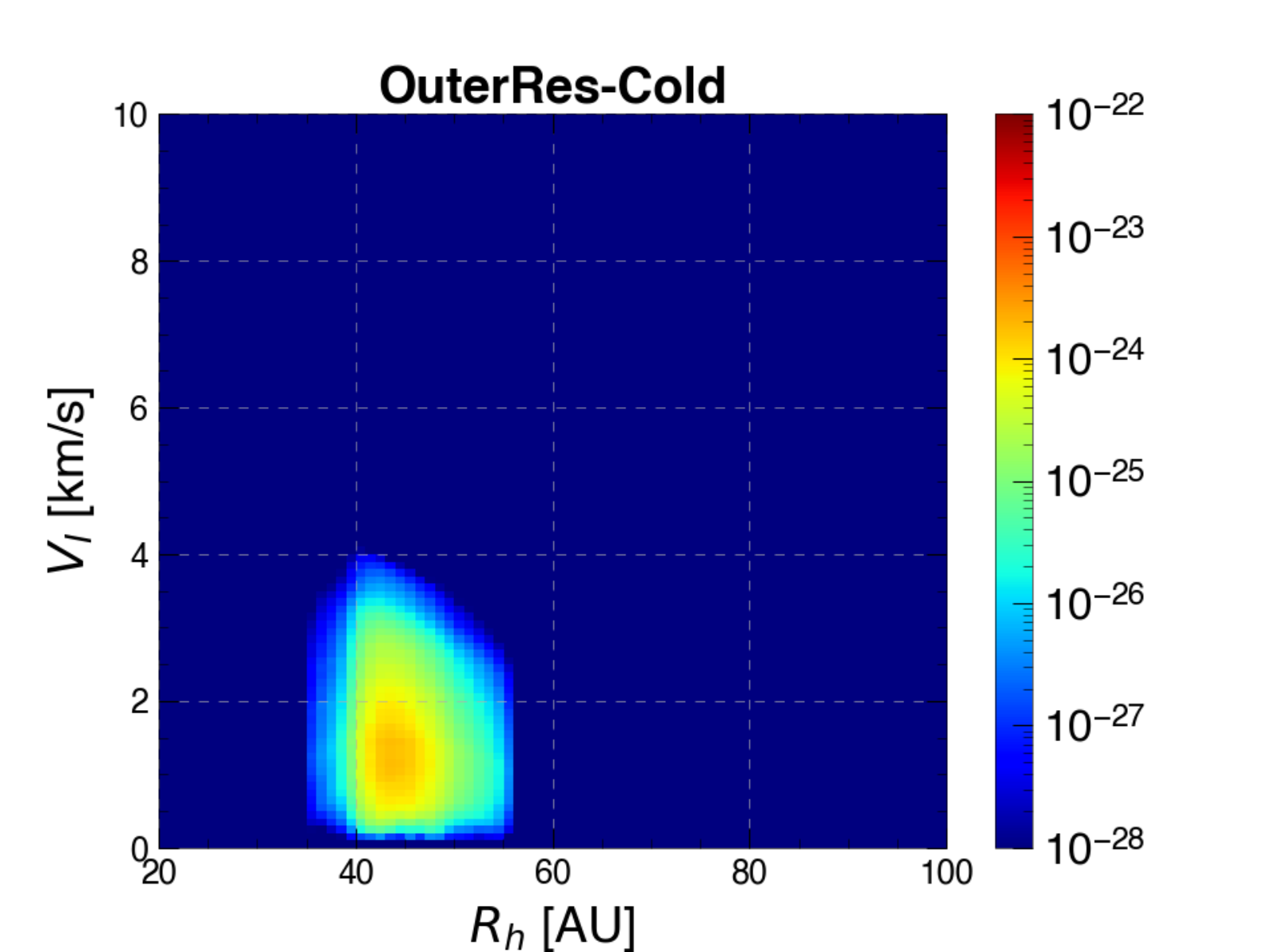}{0.33\textwidth}{(f)}}
	\gridline{\fig{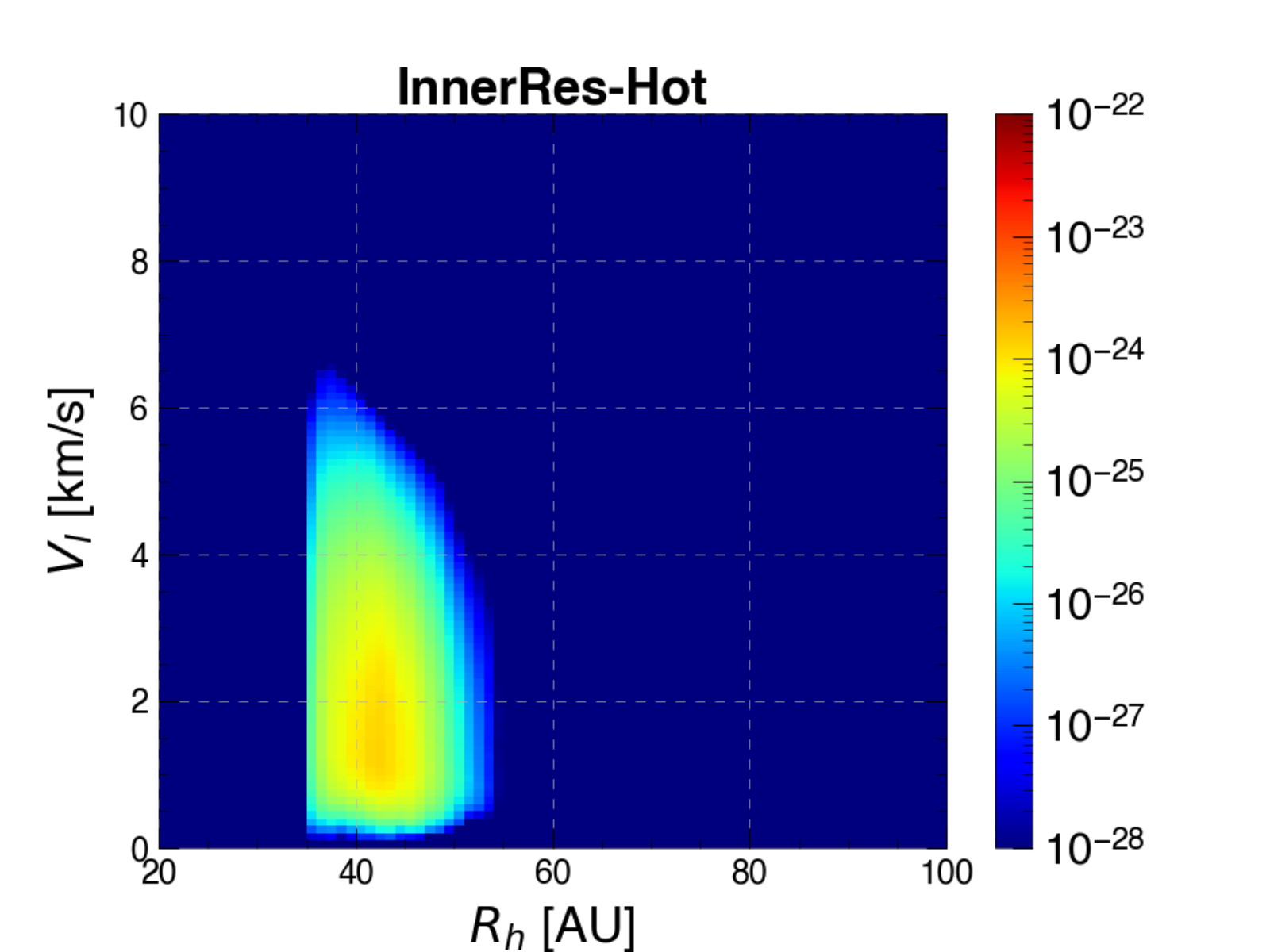}{0.33\textwidth}{(g)}
		\fig{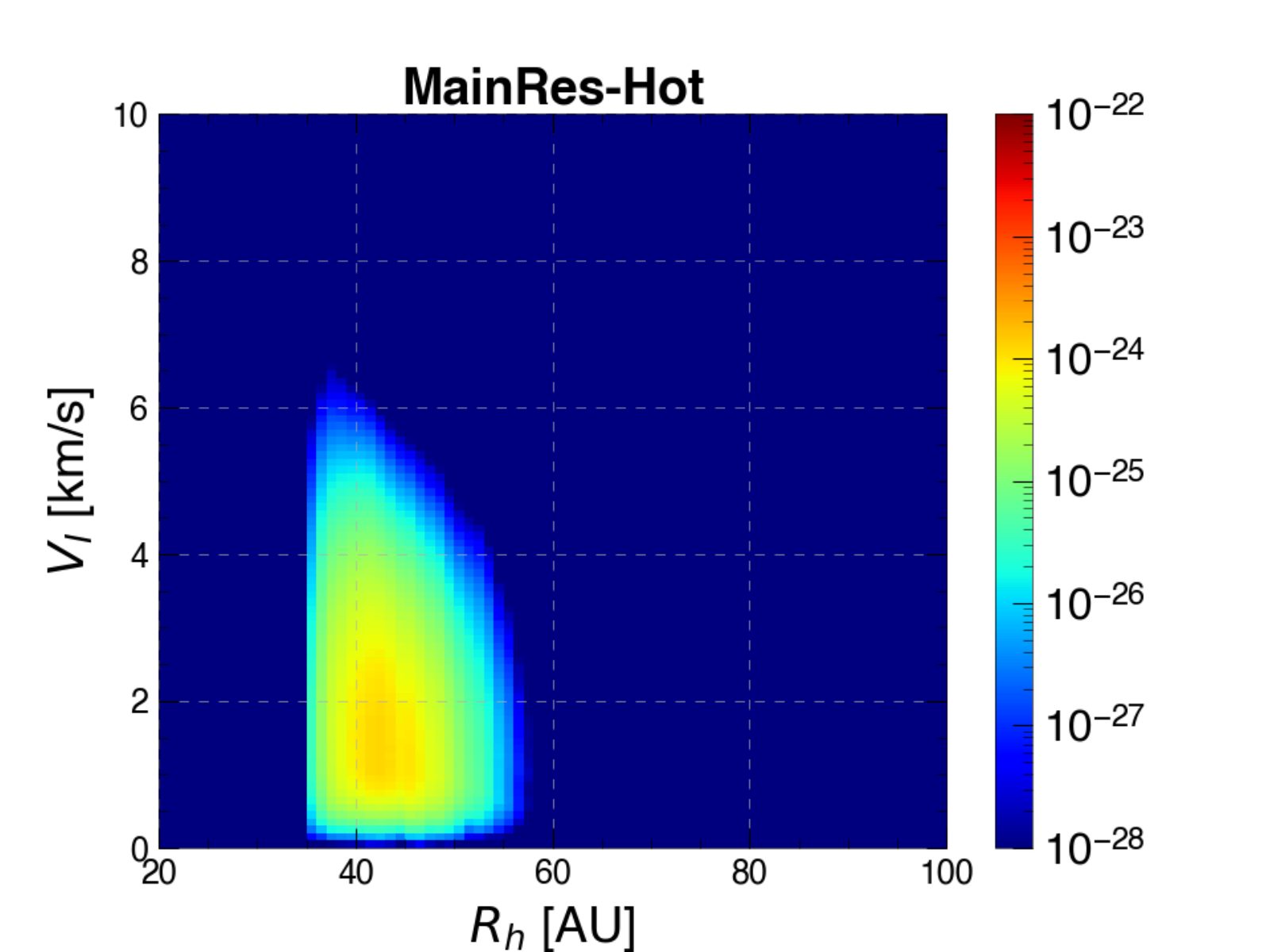}{0.33\textwidth}{(h)}
		\fig{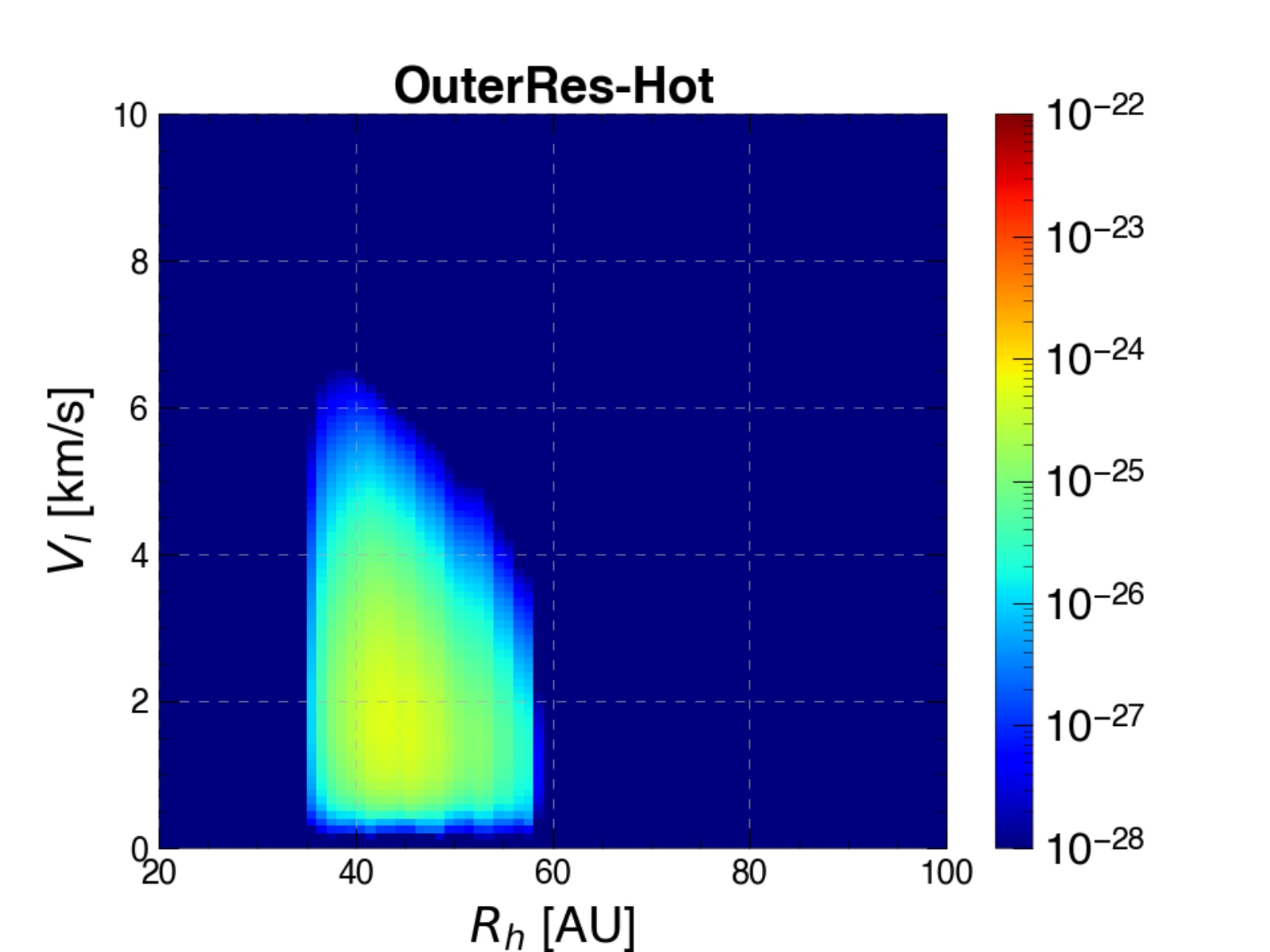}{0.33\textwidth}{(i)}}
	\caption{\footnotesize{Intrinsic collision probabilities $P_I$ (colorbar) with a given collision speed $V_I$ at a given heliocentric distance $R_h$ between different dynamical KBO classes. The intrinsic collision probability in each cell is color coded and has units km$^{-2}$~yr$^{-1}$~TNO$^{-1}$. The values of the color bars only include values $P_I > 10^{-28}$ km$^{-2}$~yr$^{-1}$, as smaller values are unlikely to result in collision over the age of the Solar System (see text for details).}}
	\label{fig:grid1}
\end{figure*}
\begin{figure*}[htpb]
	\gridline{\fig{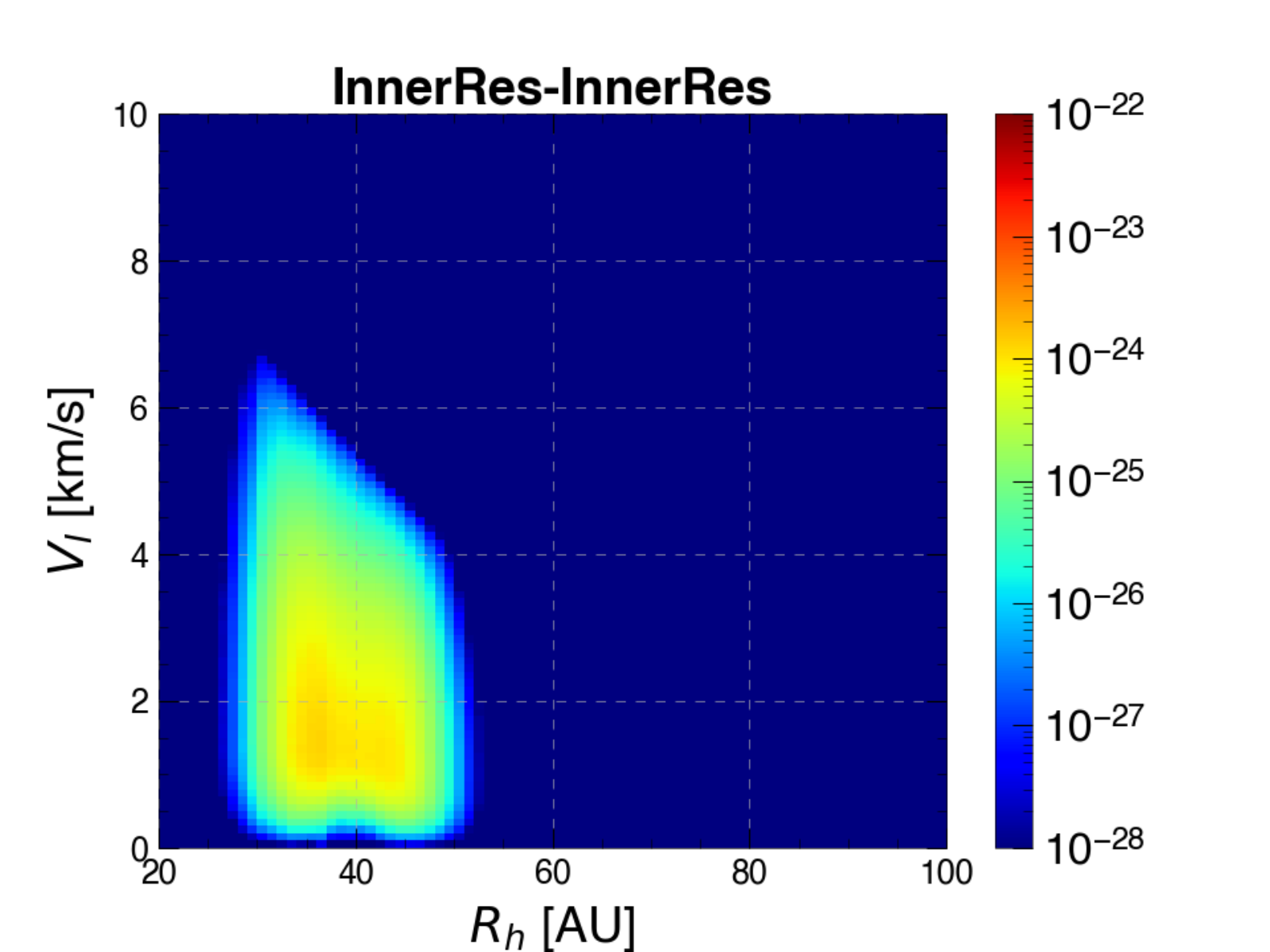}{0.33\textwidth}{(a)}
		\fig{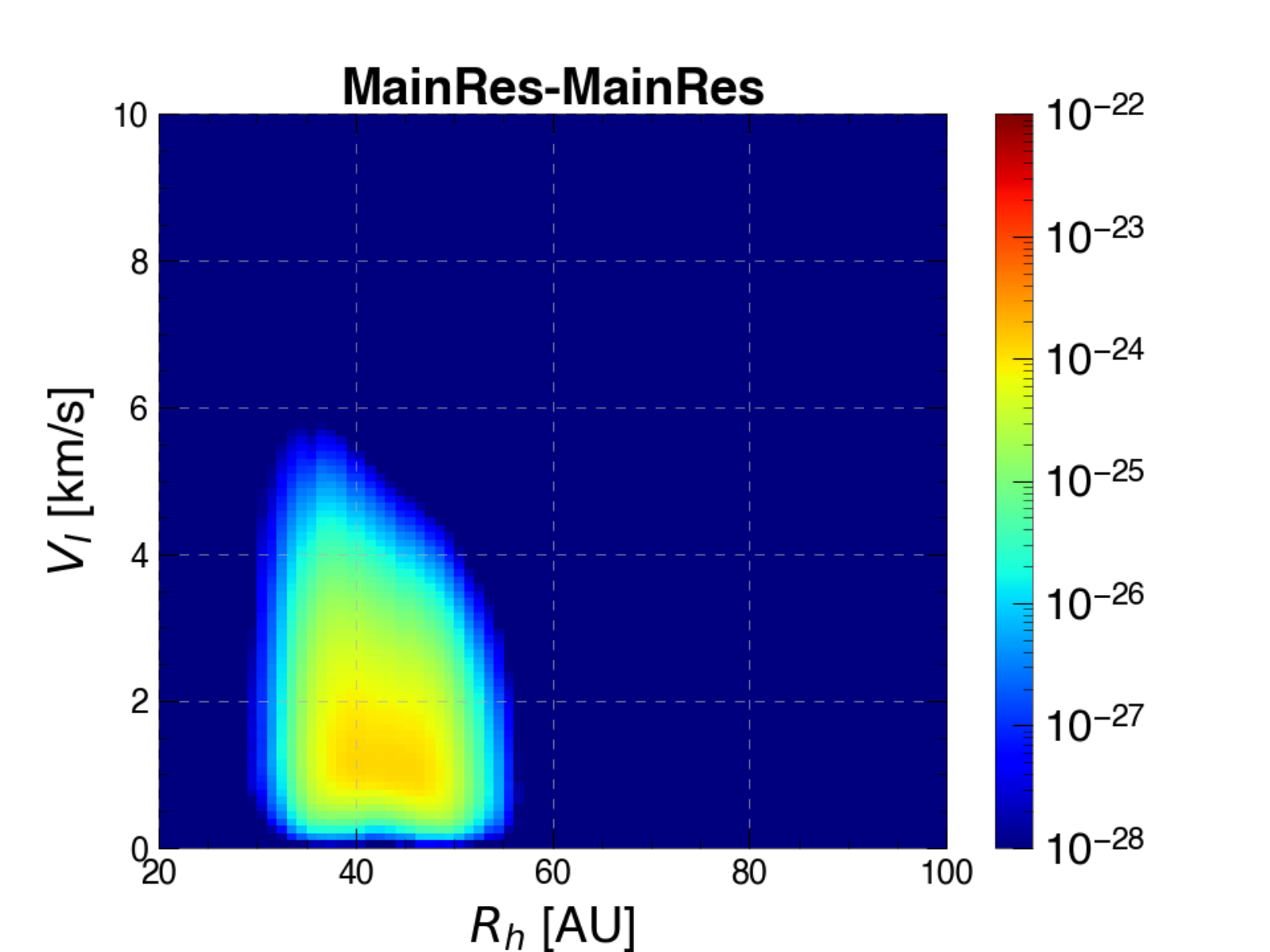}{0.33\textwidth}{(b)}
		\fig{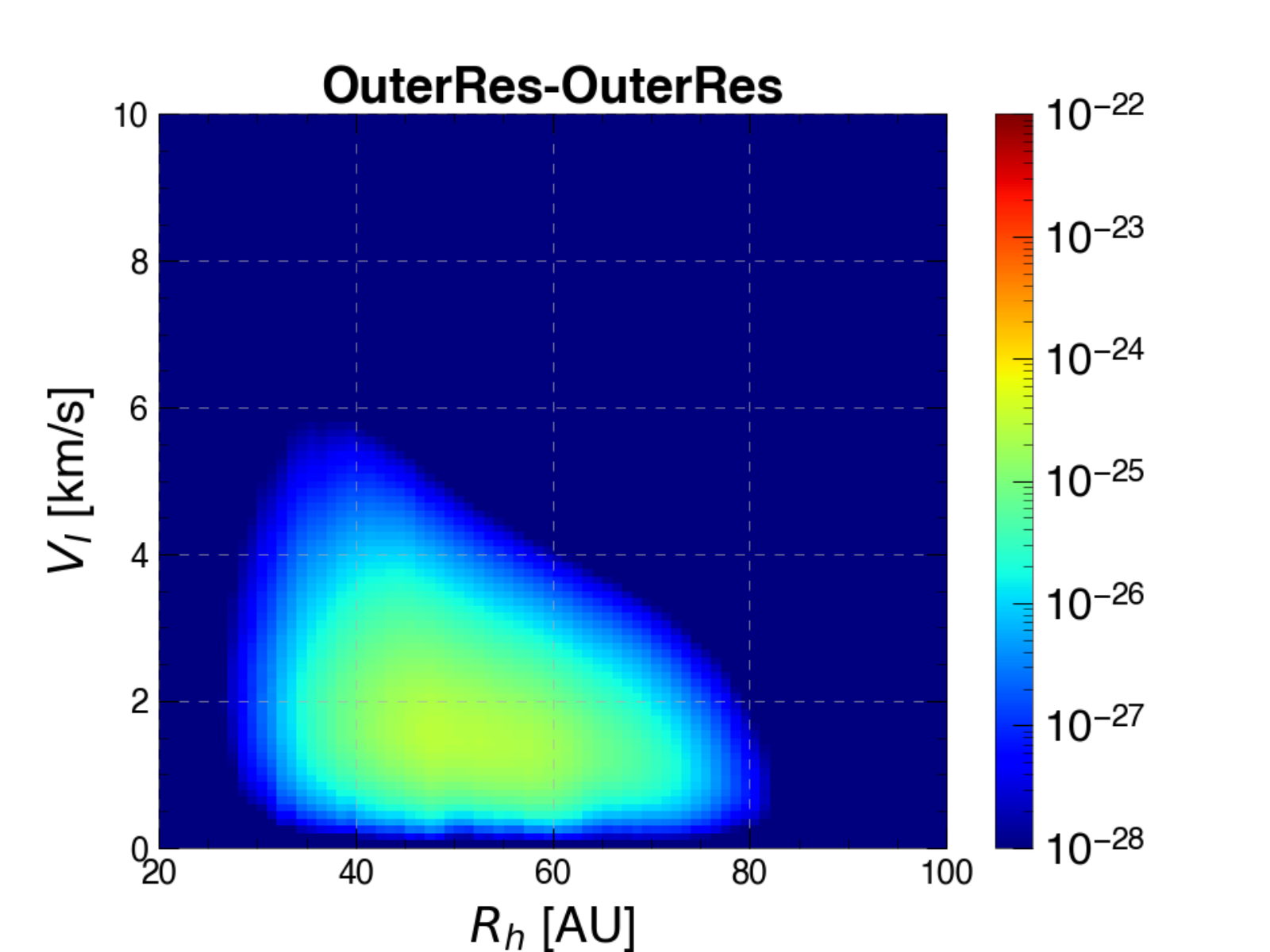}{0.33\textwidth}{(c)}}
	\gridline{\fig{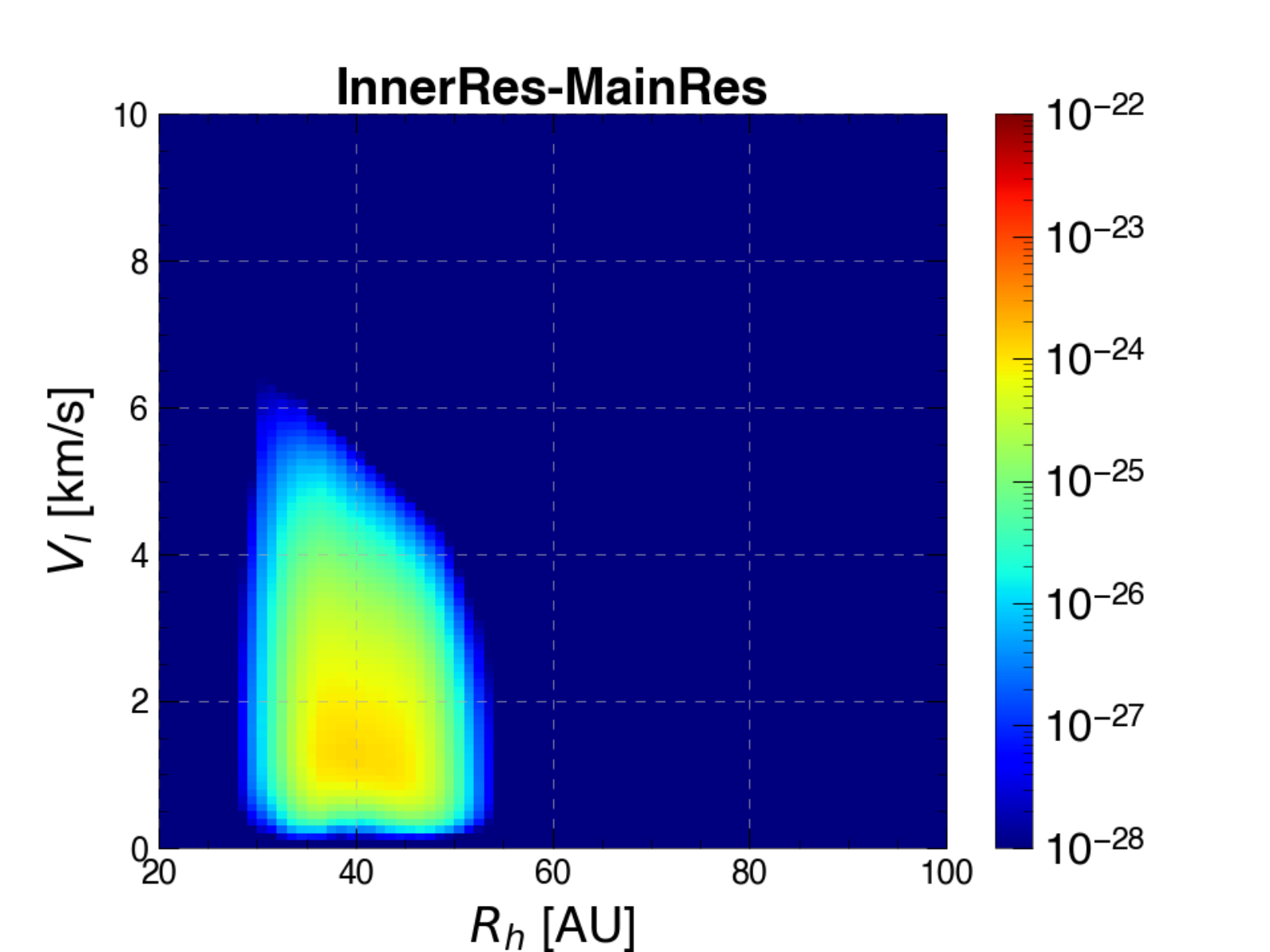}{0.33\textwidth}{(d)}
		\fig{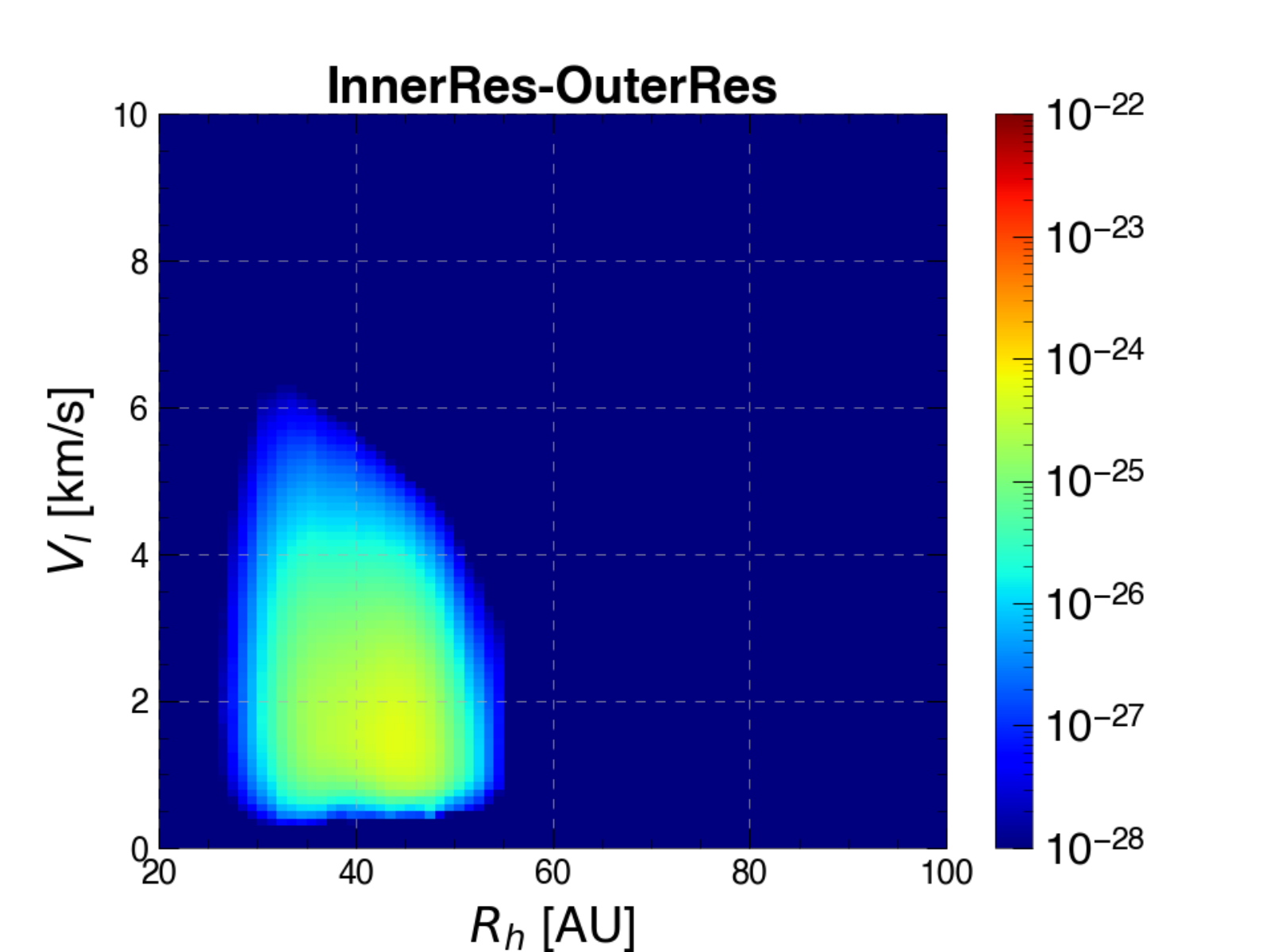}{0.33\textwidth}{(e)}
		\fig{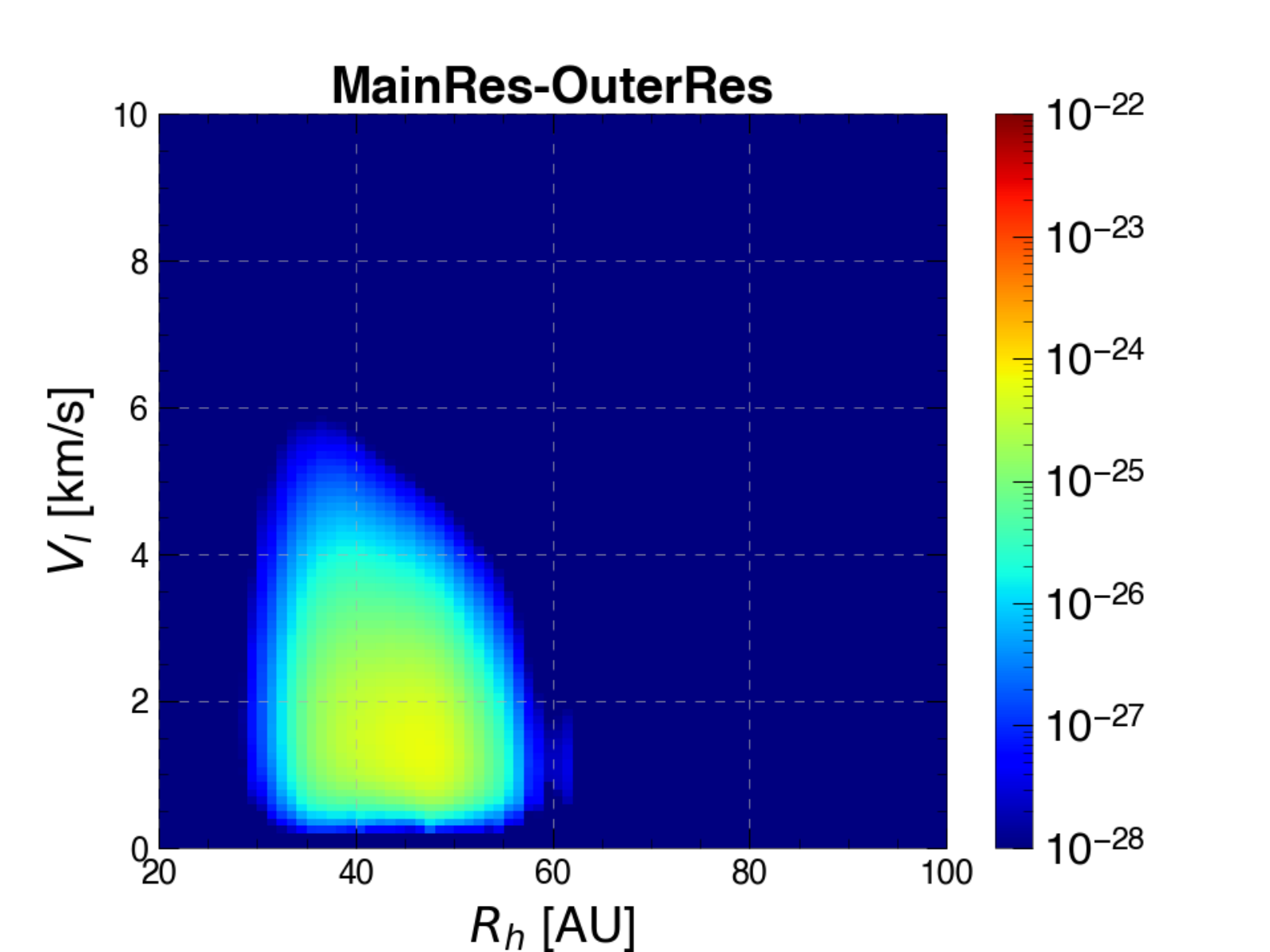}{0.33\textwidth}{(f)}}
	\gridline{\fig{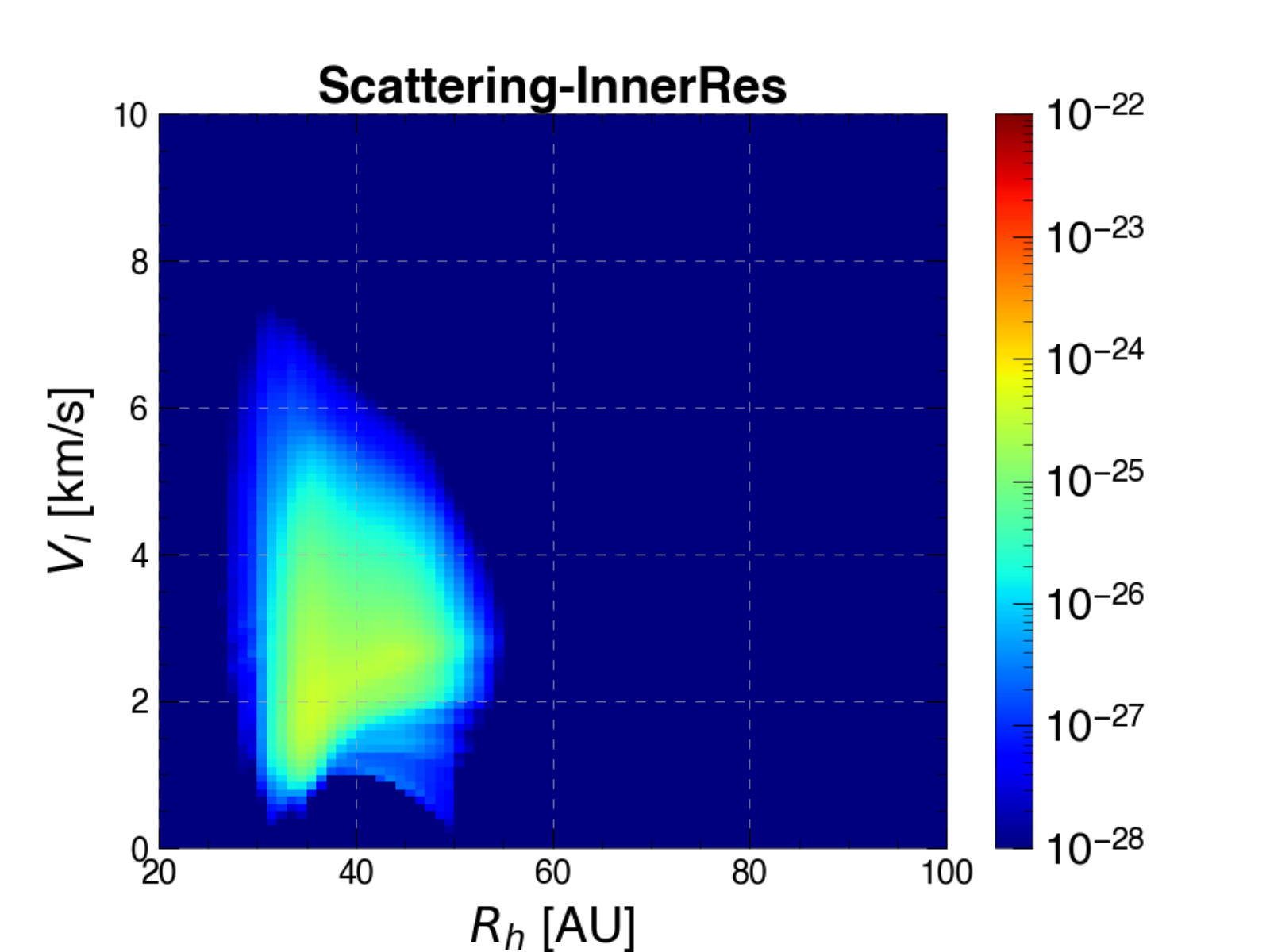}{0.33\textwidth}{(g)}
		\fig{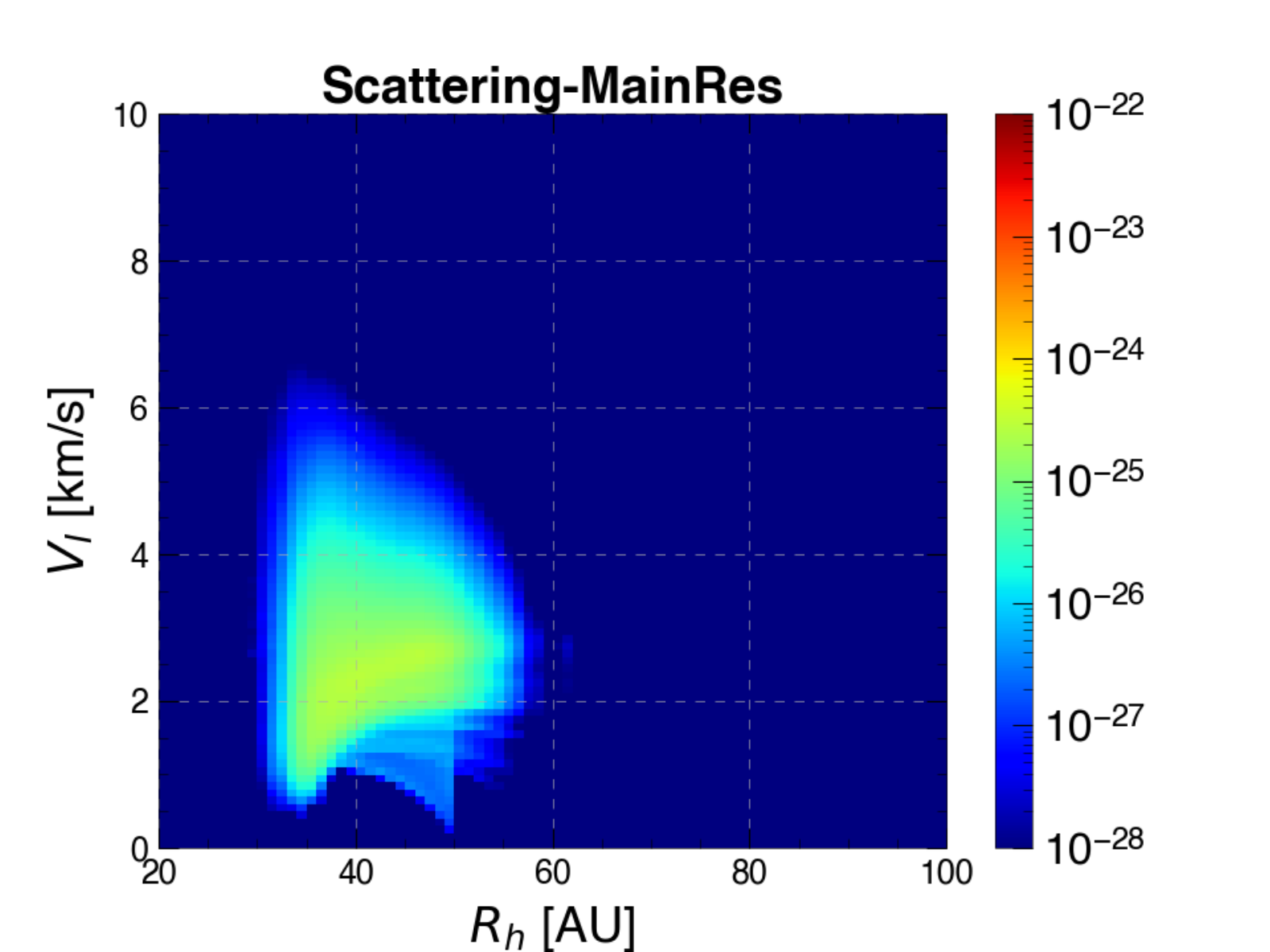}{0.33\textwidth}{(h)}
		\fig{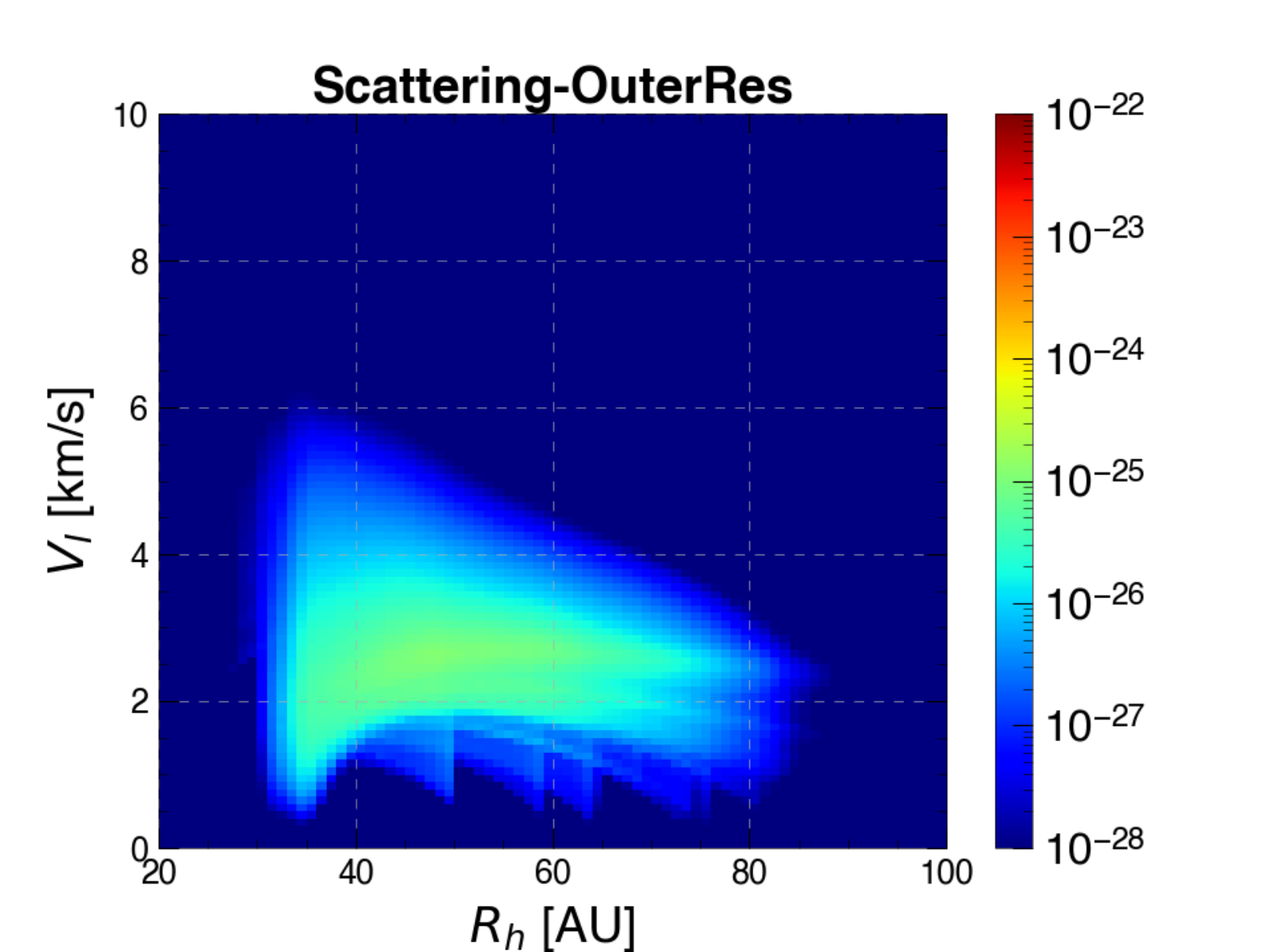}{0.33\textwidth}{(i)}}
	\caption{\footnotesize{Intrinsic collision probabilities $P_I$ (colorbar) with a given collision speed $V_I$ at a given heliocentric distance $R_h$ between different dynamical KBO classes. The intrinsic collision probability in each cell is color coded and has units km$^{-2}$~yr$^{-1}$~TNO$^{-1}$. The values of the color bars only include values $P_I > 10^{-28}$ km$^{-2}$~yr$^{-1}$, as smaller values are unlikely to result in collision over the age of the Solar System (see text for details).}}
	\label{fig:grid2}
\end{figure*}
\begin{figure*}[htpb]
	\gridline{\fig{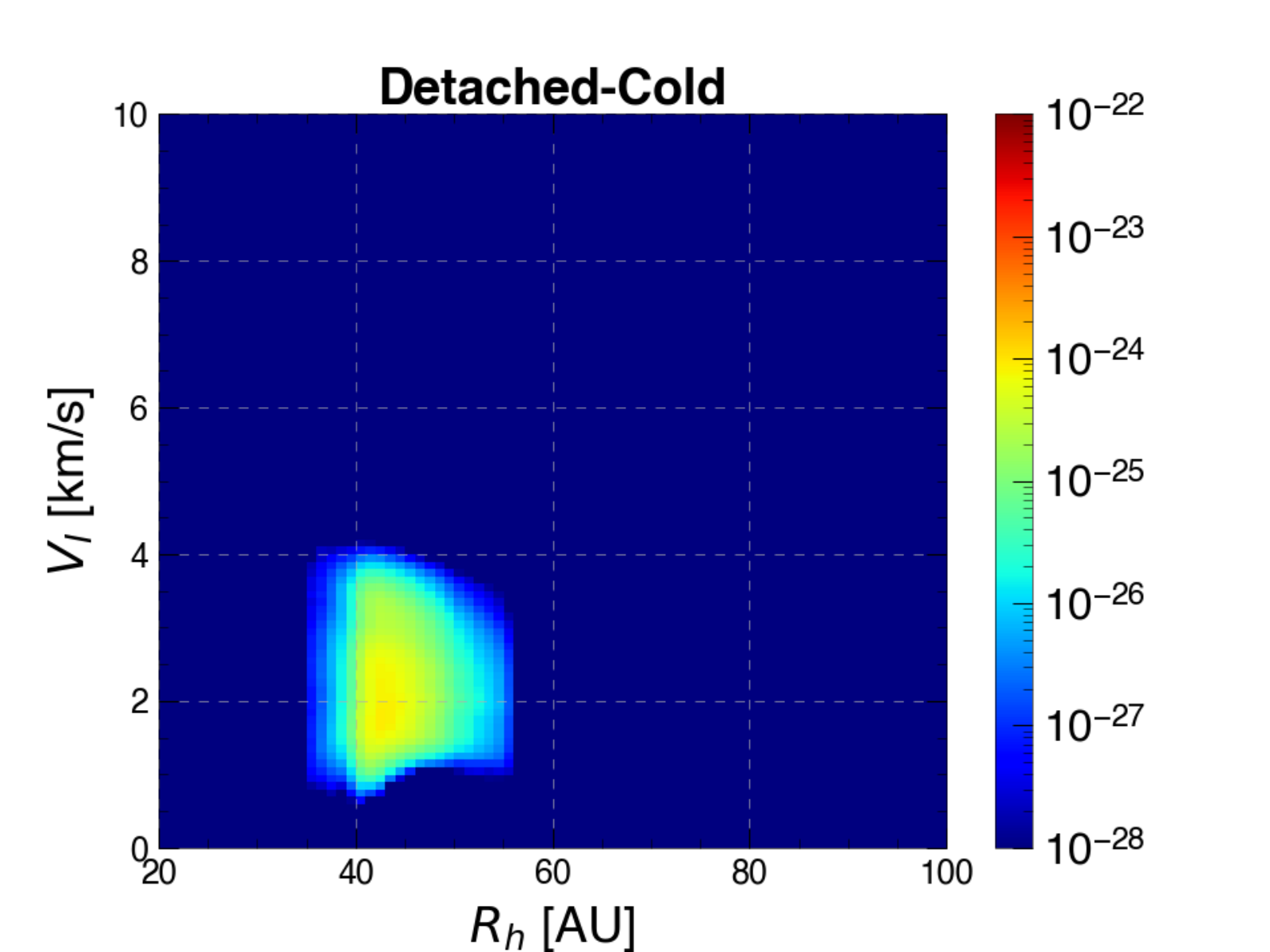}{0.33\textwidth}{(a)}
		\fig{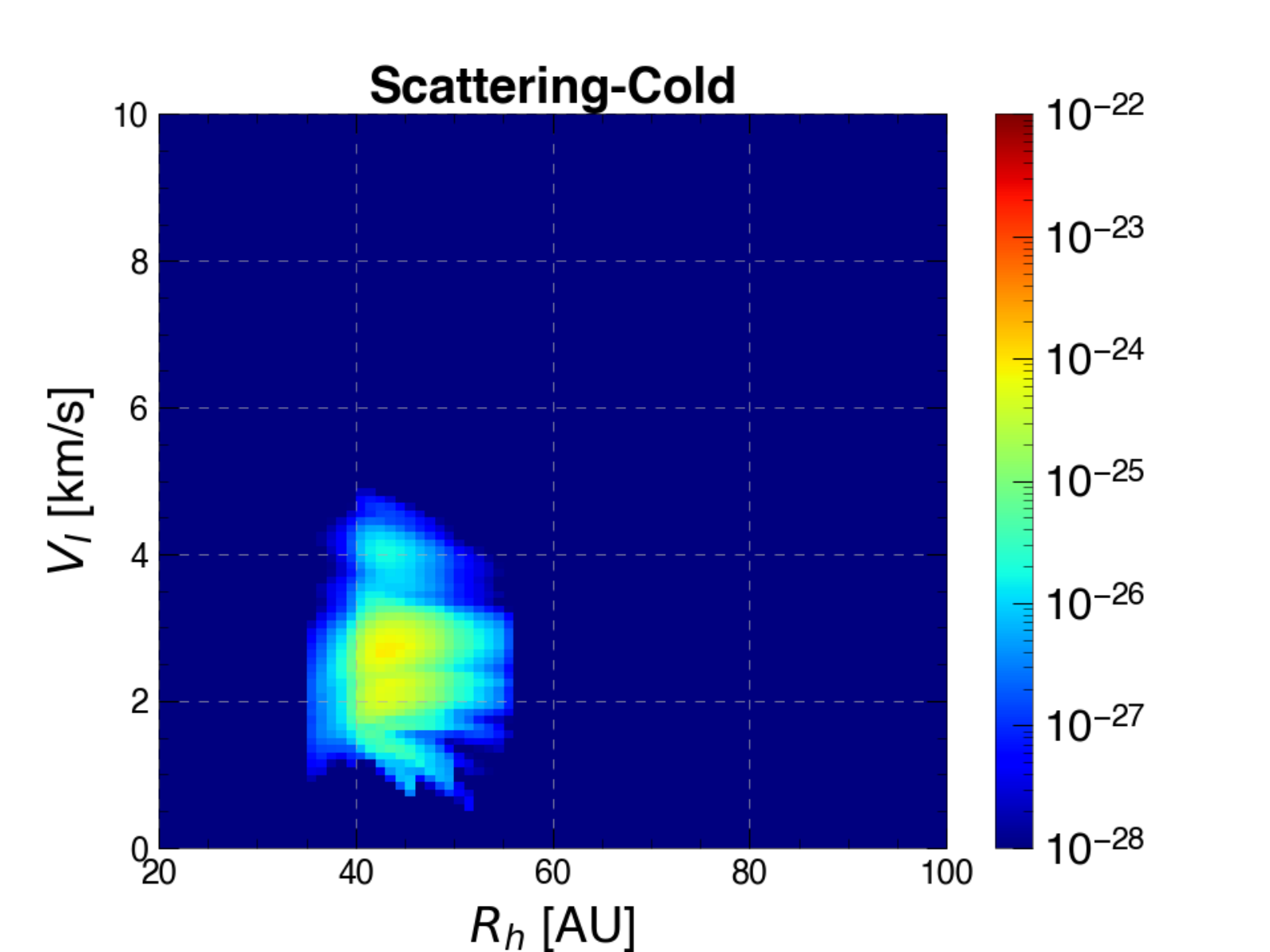}{0.33\textwidth}{(b)}
		\fig{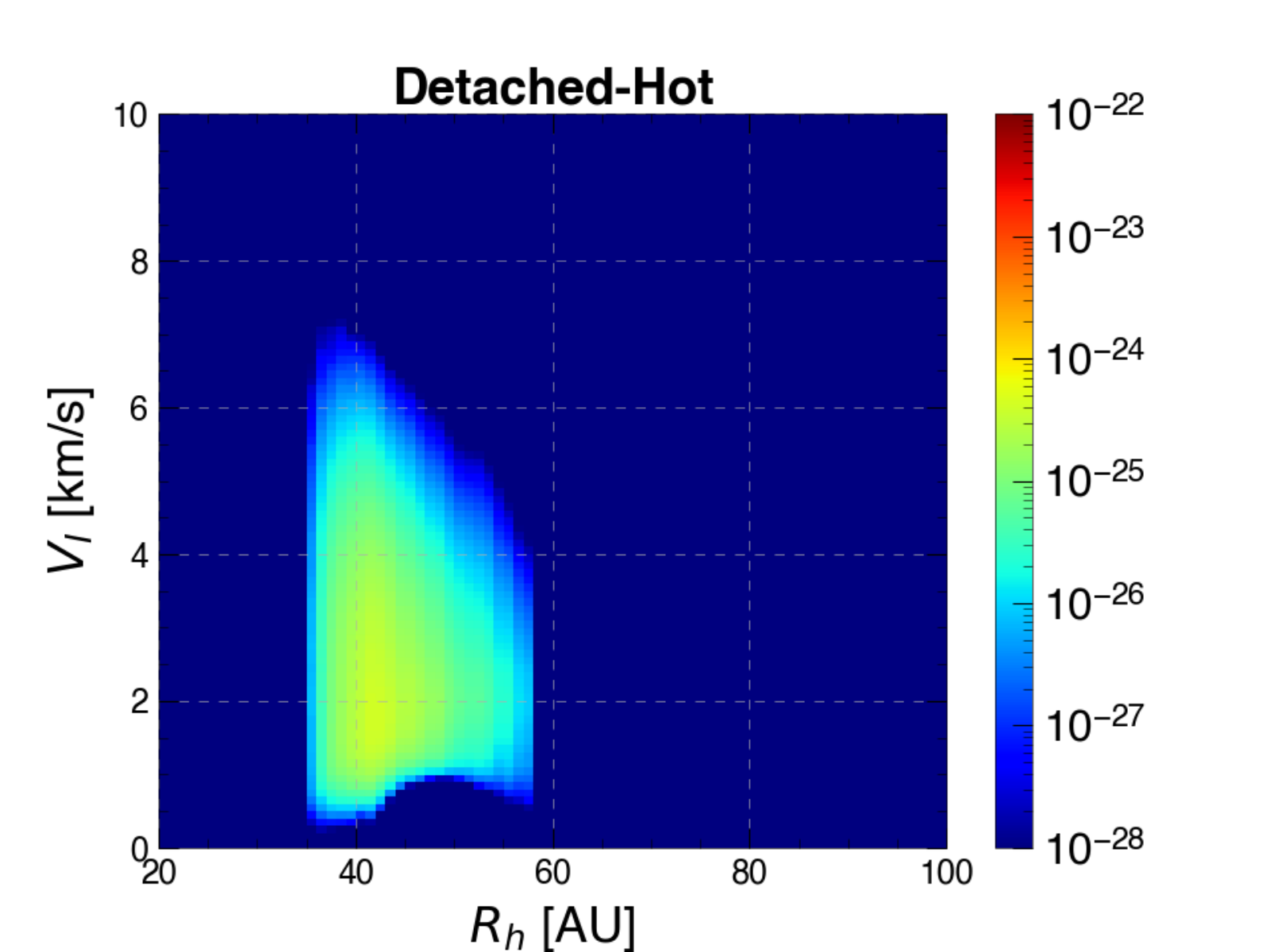}{0.33\textwidth}{(c)}}
	\gridline{\fig{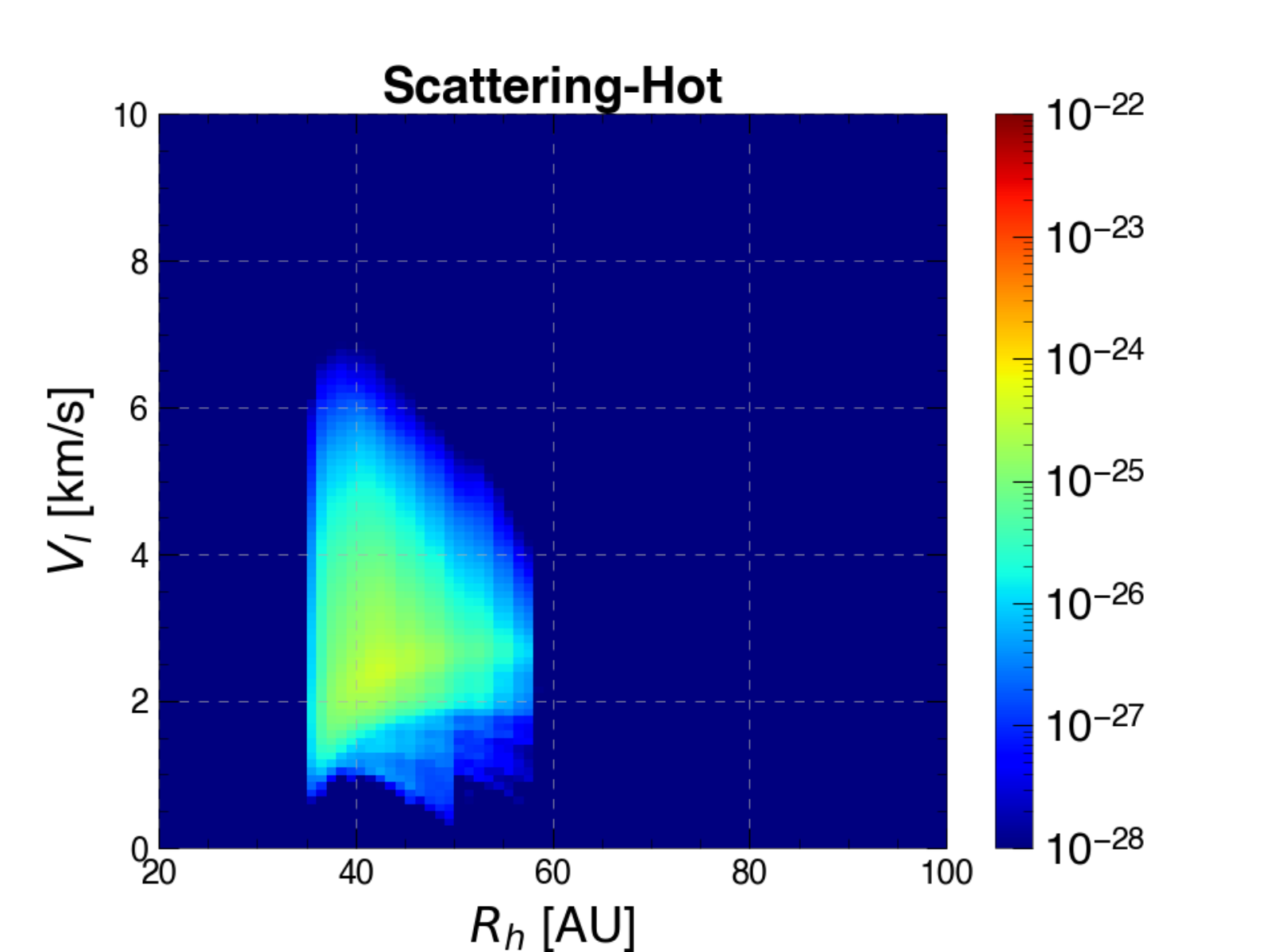}{0.33\textwidth}{(d)}
		\fig{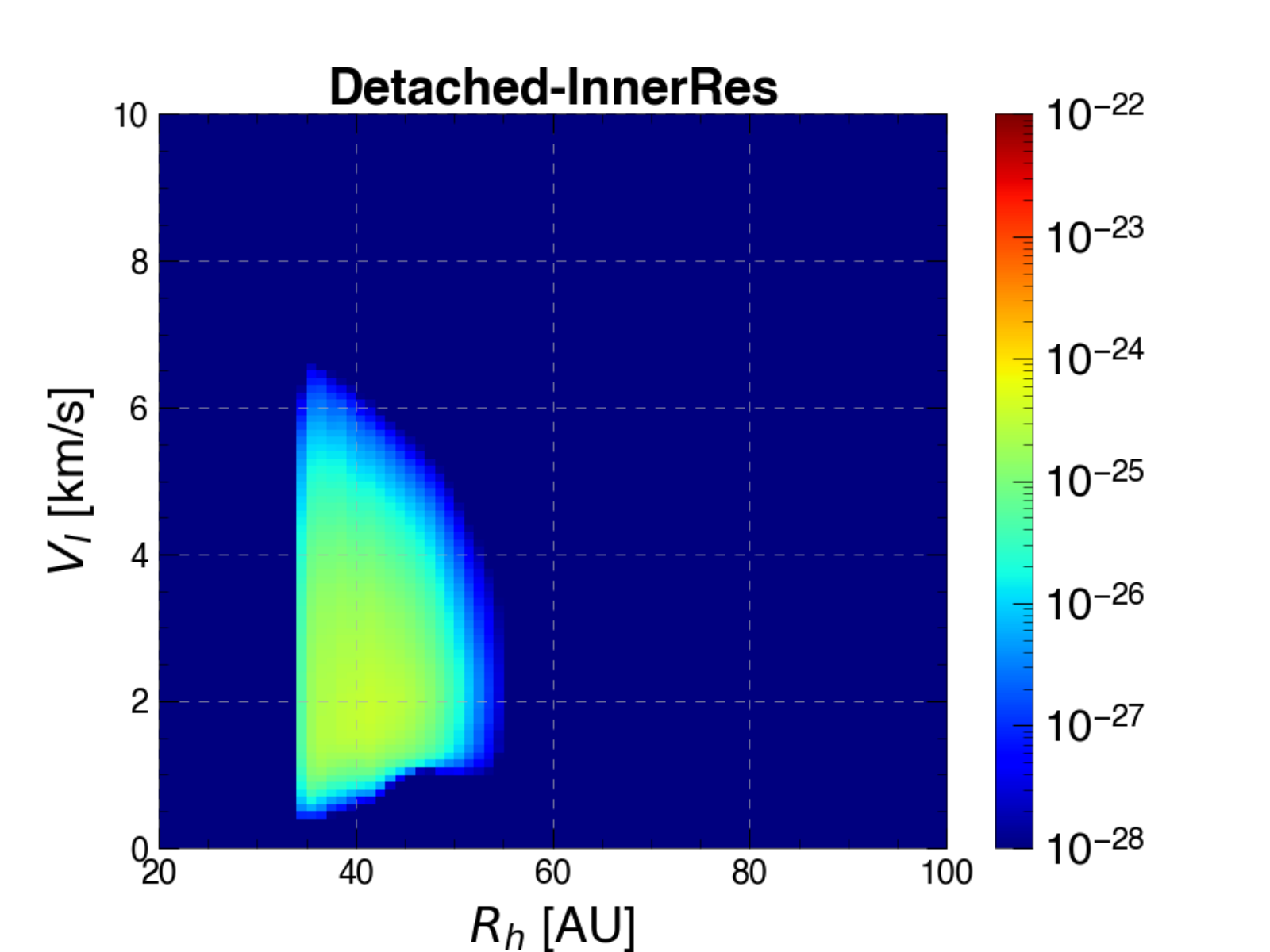}{0.33\textwidth}{(e)}
		\fig{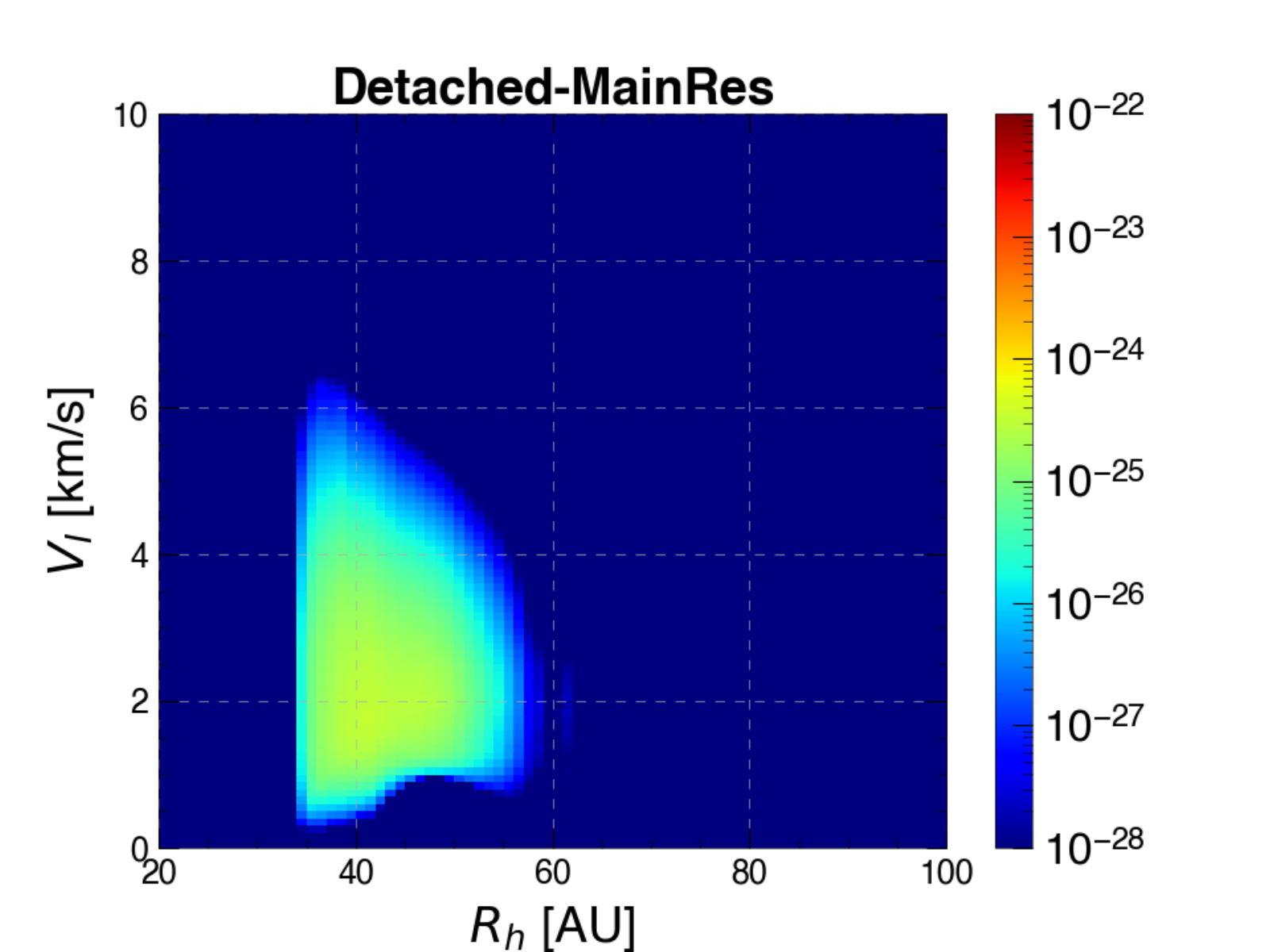}{0.33\textwidth}{(f)}}
	\gridline{\fig{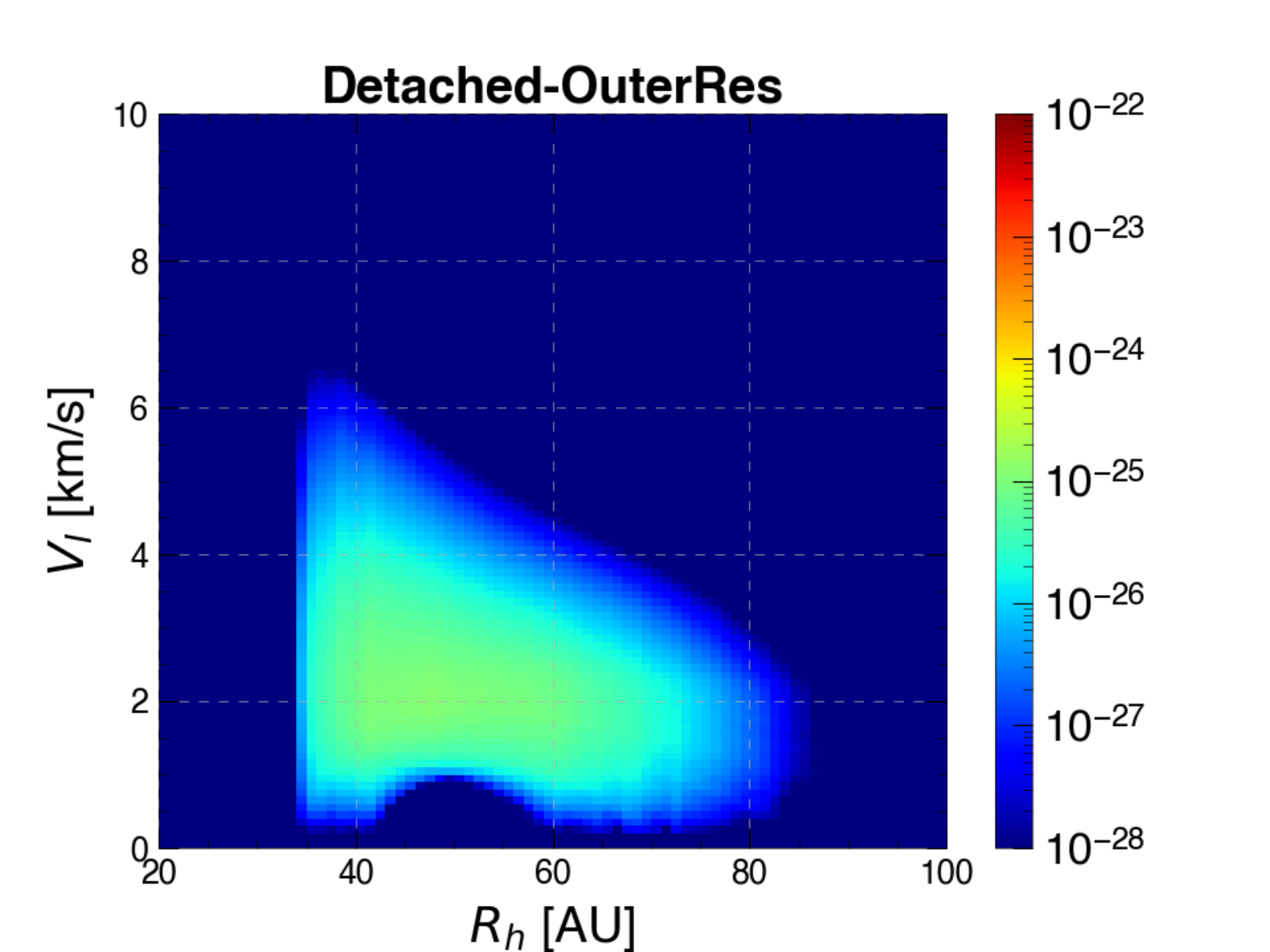}{0.33\textwidth}{(g)}
		\fig{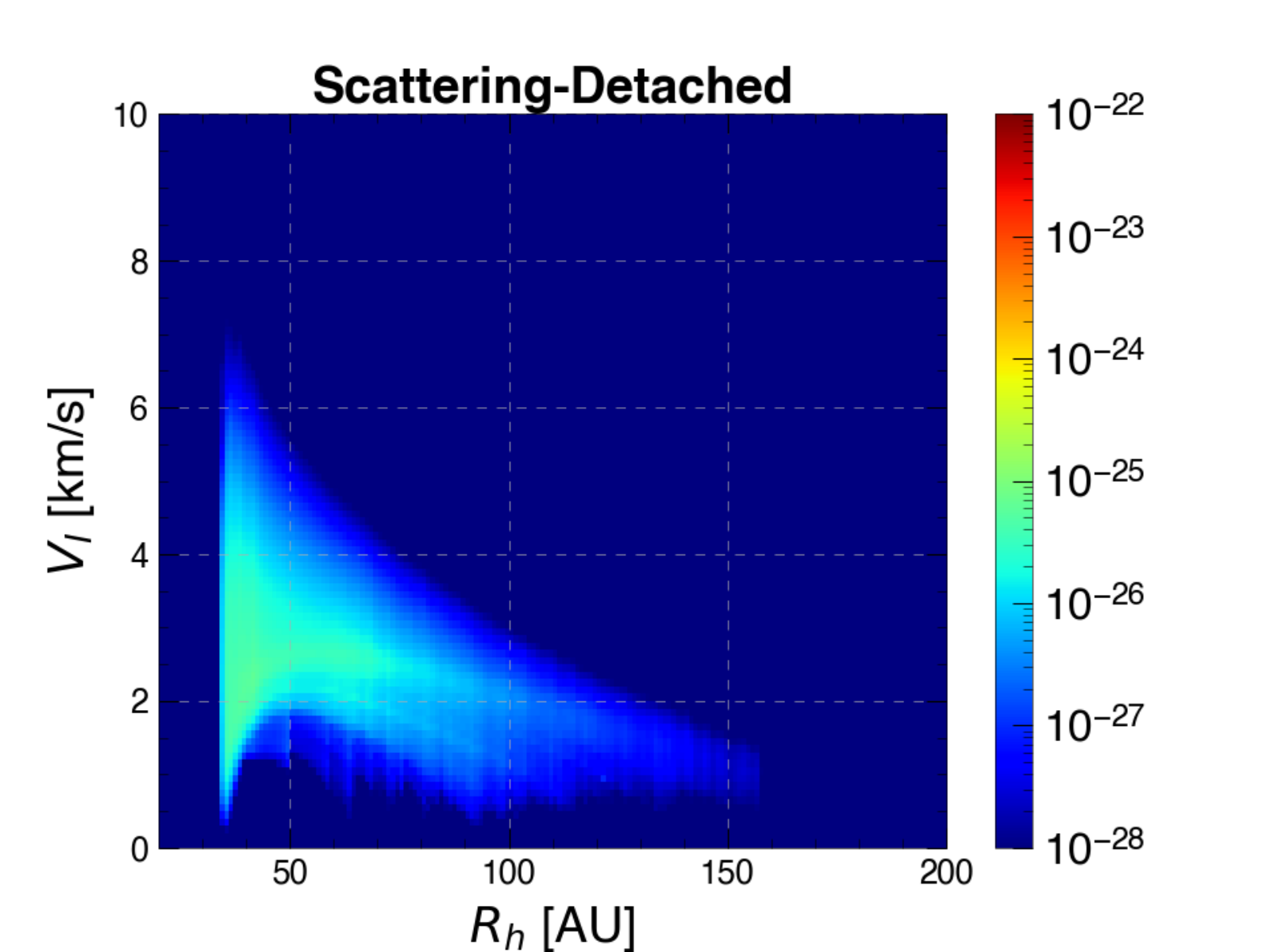}{0.33\textwidth}{(h)}
		\fig{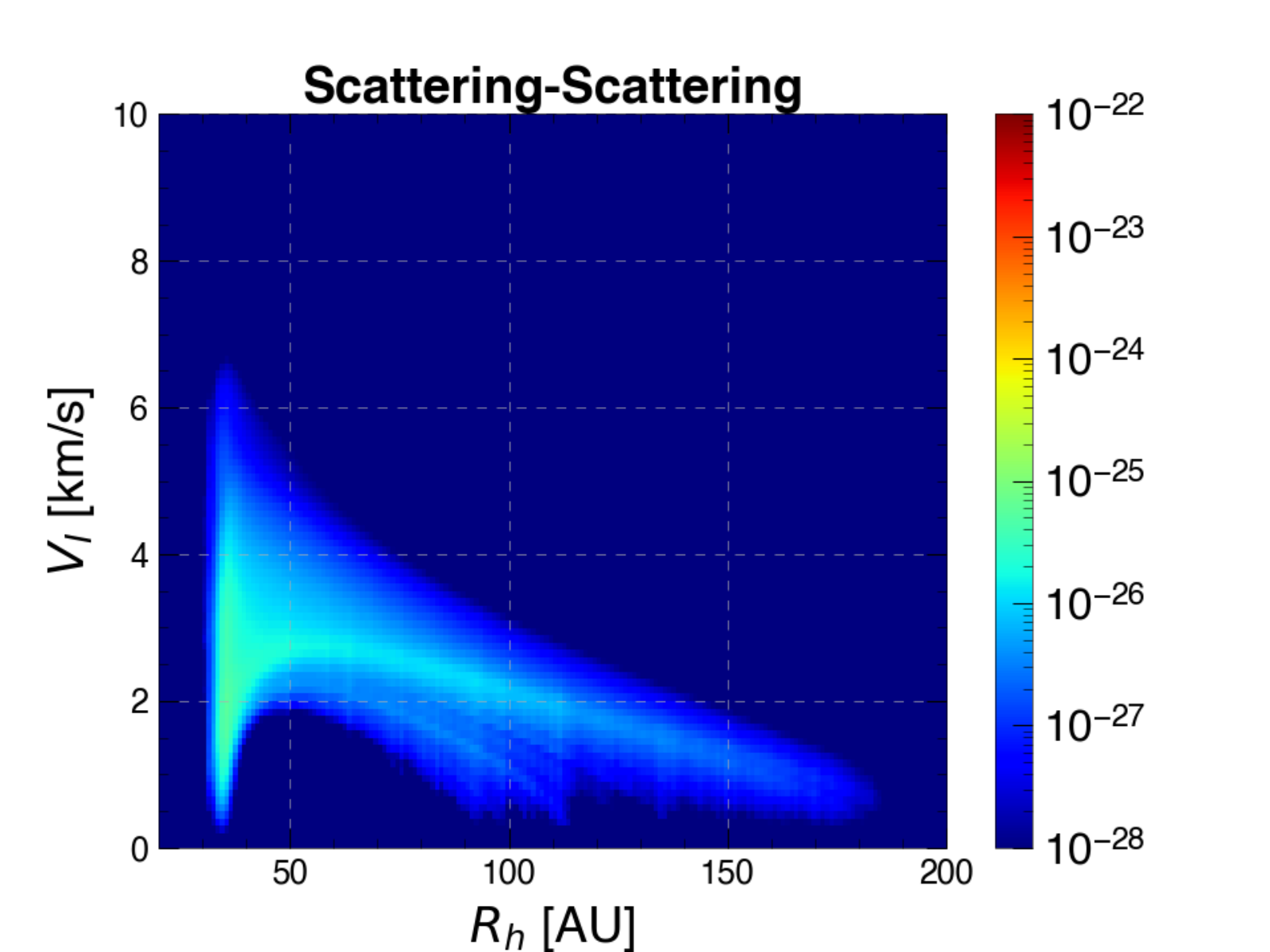}{0.33\textwidth}{(i)}}
	\gridline{\fig{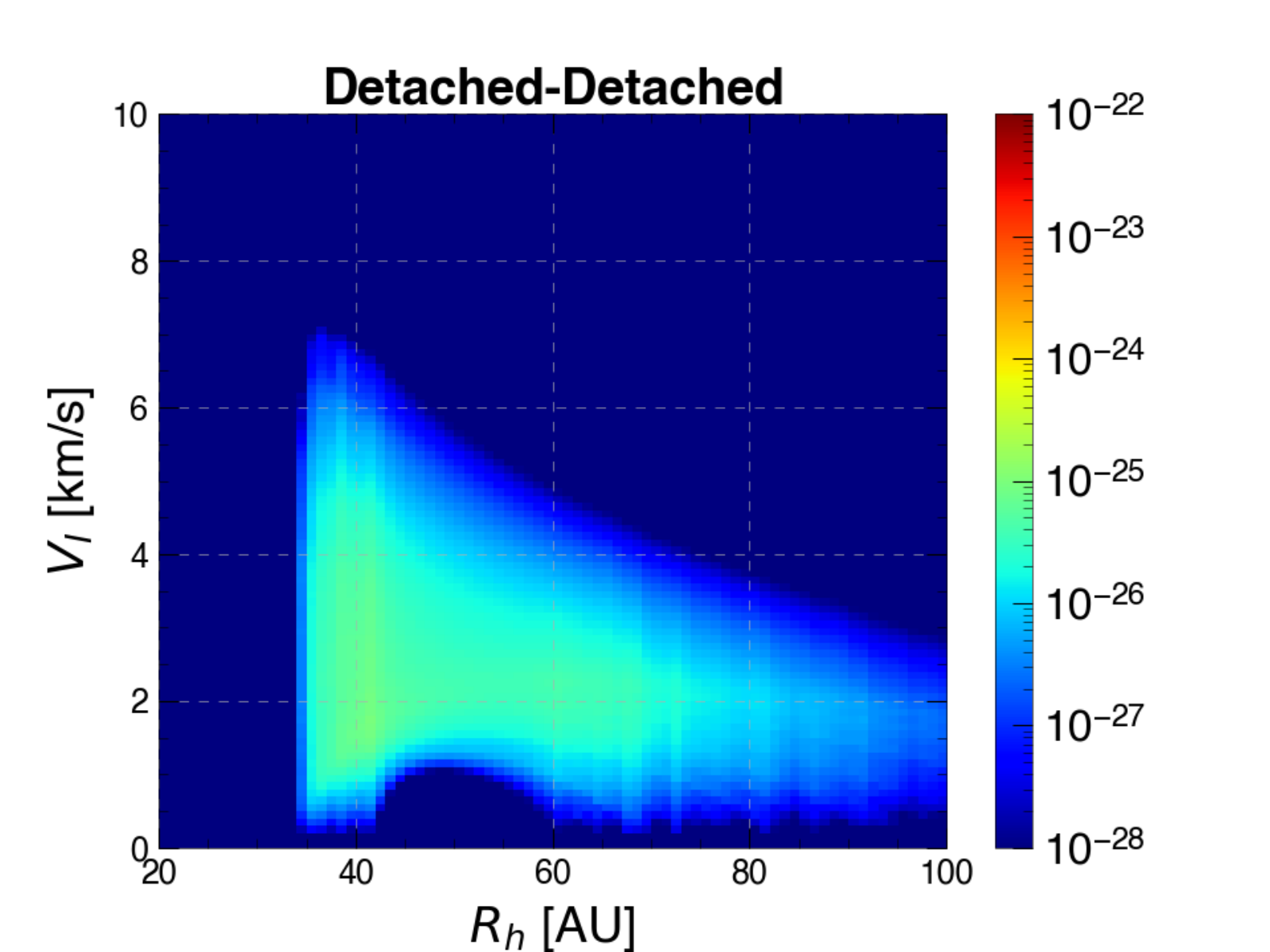}{0.33\textwidth}{(j)}
	}
	\caption{\footnotesize{Intrinsic collision probabilities $P_I$ (color bar) with a given collision speed $V_I$ at a given heliocentric distance $R_h$ between different dynamical KBO classes. The intrinsic collision probability in each cell is color coded and has units km$^{-2}$~yr$^{-1}$~TNO$^{-1}$. The values of the color bars only include values $P_I > 10^{-28}$ km$^{-2}$~yr$^{-1}$, as smaller values are unlikely to result in collision over the age of the Solar System (see text for details).}}
	\label{fig:grid3}
	\vspace{-50.69pt}
\end{figure*}

The results of the calculated intrinsic collision probability $P_I$ and speeds $V_I$, between and within different KBO populations, are summarized in Figures \ref{fig:grid1} \ref{fig:grid2}, \ref{fig:grid3}. Each pixel in each subplot is color coded in terms of the intrinsic collision probability $P_I$ in [km$^{-2}$~yr$^{-1}$TNO$^{-1}$] occurring with a collision speed $V_I$ in [km/s] at a given heliocentric distance $R_h$ in [AU]. Given the large range of variation of $\Delta P_I$ (over seven orders of magnitude), the data in each plot is presented in a logarithmic scale.  
In addition, we also provide separately the probability density of a collision happening with a given impact speed $V_I$, for the different dynamical TNO populations (Fig.~\ref{figs:ispeeds}).  
The intrinsic collision probabilities between each KBO population i.e., the average of all pixel values for each subplot are summarized in Table\ref{Tab:cols}. 
The intrinsic probabilities must be multiplied by the size of the populations to determine the collision rate. 
The probability plots provide a quick visualization of where in the Kuiper belt collisions are most likely to occur and which orbital classes have the largest intrinsic probability of collisions.


Our calculated intrinsic probabilities and speeds indicate that collisions in the Kuiper belt are possible over a wide range of the heliocentric distances $R_h=30-200$~AU and speeds $V_I<9$~km/s, though the majority of these collisions are expected to occur in a narrow range between $R_h = 30-55$~AU and with much lower speeds (see Fig.\ref{fig:grid1}, \ref{fig:grid2} and \ref{fig:grid3}). Although, we see collision speeds as high as 9 km/s, there are very unlikely to occur over the age of the Solar System with speed of $V_I <6$~km/s being more likely (Fig.~\ref{figs:ispeeds}). 

The highest intrinsic collision probability occurs within the ``cold" sub-component of the classical belt, followed by the ``hot" on ``cold" impacts, and within the ``hot" substructure. 
This is not surprising given that these populations are more concentrated in the orbit phase-space compared to the other orbit classes considered. 
It is also evident that the cold classicals collide with median velocity of $\langle V_I\rangle \approx 0.3$ km/s and have a narrow impact speed range (see Fig.\ref{figs:ispeeds} (a)). 
That raises an interesting question as to whether the collisions between these objects are disruptive,  crater forming or result in complex structure formation. 
The expectation of low impact velocities is similar to the result reported in \citet{green19} when considering impacts onto Arrokoth and utilizing the CFEPS L7 orbit model.  

We also note a bi-modality in the impact speed distribution for ``hot" on ``cold" collisions (Fig~\ref{figs:ispeeds}a). The reason for this is the existence of substructures in the semi-major axis distribution of the ``cold" and ``hot" sub-components of the main belt creating distinct sub-sets of collisional circumstances.
A bi-modal velocity distribution is also seen for the scattering population colliding with the ``cold" sub-structure of the main belt (Fig.\ref{figs:ispeeds}d).
This structure also appears to be due to the sub-structure in the semi-major axis distribution of the ``cold" population. 
 
\begin{deluxetable*}{r@{ - }llccccc}
\tablecaption{Intrinsic collision probability $ \langle P_I \rangle $ and speeds $V_I$ by sub-component.
\label{Tab:cols}} 
\tablehead{
\multicolumn2c{Population} & 
\colhead{$\langle P_I \rangle$} & 
\colhead{Relative $\langle P_I \rangle$\tablenotemark{a}} &
\multicolumn3c{$V_I (\text{km~s}^{-1})$} \\
Target & Impactor
& \colhead{ km$^{-2}$yr$^{-1}$ } 
& \colhead{\%}
& \colhead{0.05} & \colhead{0.50} & \colhead{0.95}
}
\startdata
\sidehead{Cold Classical}
Cold & Cold  & 7.5e-22       &16.6& 0.1 & 0.3 &0.7\\
Cold & Hot&  3.2e-22         &7.1& 0.6 & 1.4 & 3.0  \\
Cold & Inner res & 2.6e-22  &5.8&  0.5  & 1.3 &2.7  \\
Cold & Main res  & 3.0e-22  &6.7&  0.4  & 1.0 &2.1   \\ 
Cold & Outer res & 1.6e-22  &3.6&  0.5  & 1.2 & 2.5   \\
Cold & Detached &   7.7e-23  &1.7& 1.3 & 2.1 & 3.4 \\
Cold & Scattering &  4.7e-23   &1.0& 1.8  & 2.6 & 3.5   \\
\sidehead{Hot Classical}
Hot & Hot & 2.9e-22          &6.4& 0.6 & 2.0 & 4.4  \\
Hot & Inner res & 2.3e-22    &5.1& 0.5 & 2.5 & 3.7  \\
Hot & Main res &  2.5e-22   &5.5&  0.6 & 1.8 & 3.8   \\ 
Hot & Outer res &   1.1e-22 &2.4& 0.7 &1.8 & 3.8 \\
Hot & Detached &   8.3e-23   &1.8& 1.1 & 2.4 & 4.7  \\
Hot & Scattering &   4.4e-23   &1.0& 1.8 & 2.7 & 4.5  \\
\sidehead{Resonant}
Inner res & Inner Res. &  3.2e-22  &7.1& 0.8 & 2.0 &4 .3   \\
Inner res & Main Res. &  2.6e-22  &5.8& 0.7 & 1.7 & 3.7 \\ 
Inner res & Outer Res. & 1.0e-22  &2.2& 0.9 & 1.9 & 3.9  \\
Main res & Main Res. &   2.6e-22   &5.8& 0.6 & 1.5 & 3.1 \\
Main res & Outer Res. & 1.2e-22    &2.7&  0.8 & 1.7 & 3.2 \\
Outer res & Outer Res. &  9.2e-23 &2.0&  0.7 & 1.6 & 3.1 \\
\sidehead{Scattering}
Scattering & Inner Res. &  6.1e-23 &1.4& 1.6 & 2.8 & 4.7  \\
Scattering & Main Res. &  5.5e-23 &1.2&  1.6 & 2.7 & 4.1 \\ 
Scattering & Outer Res. & 3.3e-23 &0.7&  1.7 & 2.8 & 4.1  \\
Scattering & Detached  & 2.3e-23  &0.5&  1.5 &  2.6 &  4.0\\
Scattering & Scattering & 1.6e-23 &0.4&  1.7 &  2.8 &  4.5 \\ 
\sidehead{Detached/Outer}
Detached & Inner Res.&  7.8e-23  &1.7& 1.2 & 2.2 & 3.9 \\
Detached & Main Res. &   8.2e-23  &1.8& 1.2 & 2.2 & 4.0  \\ 
Detached & Outer Res. & 5.2e-23  &1.2& 1.1 & 2.2 & 4.0  \\
Detached & Detached   & 3.8e-23   &0.8&  1.1 &  2.2 &  4.4 \\
\enddata
\tablenotetext{a}{Fraction of total intrinsic probability, not including relative population sizes.}
\end{deluxetable*}

The intrinsic probabilities we report here are consistent (within an order of magnitude) with those reported in \citet{delloro13} who performed similar analysis using the CFEPS-L7 model.  The impact speeds reported in \citet{delloro13}, however, are somewhat ($\sim 70\%$) higher compared to the values reported here and in \citet{green19} (which were also derived using the CFEPS L7 model).


\begin{figure*}[htpb]
	\gridline{\fig{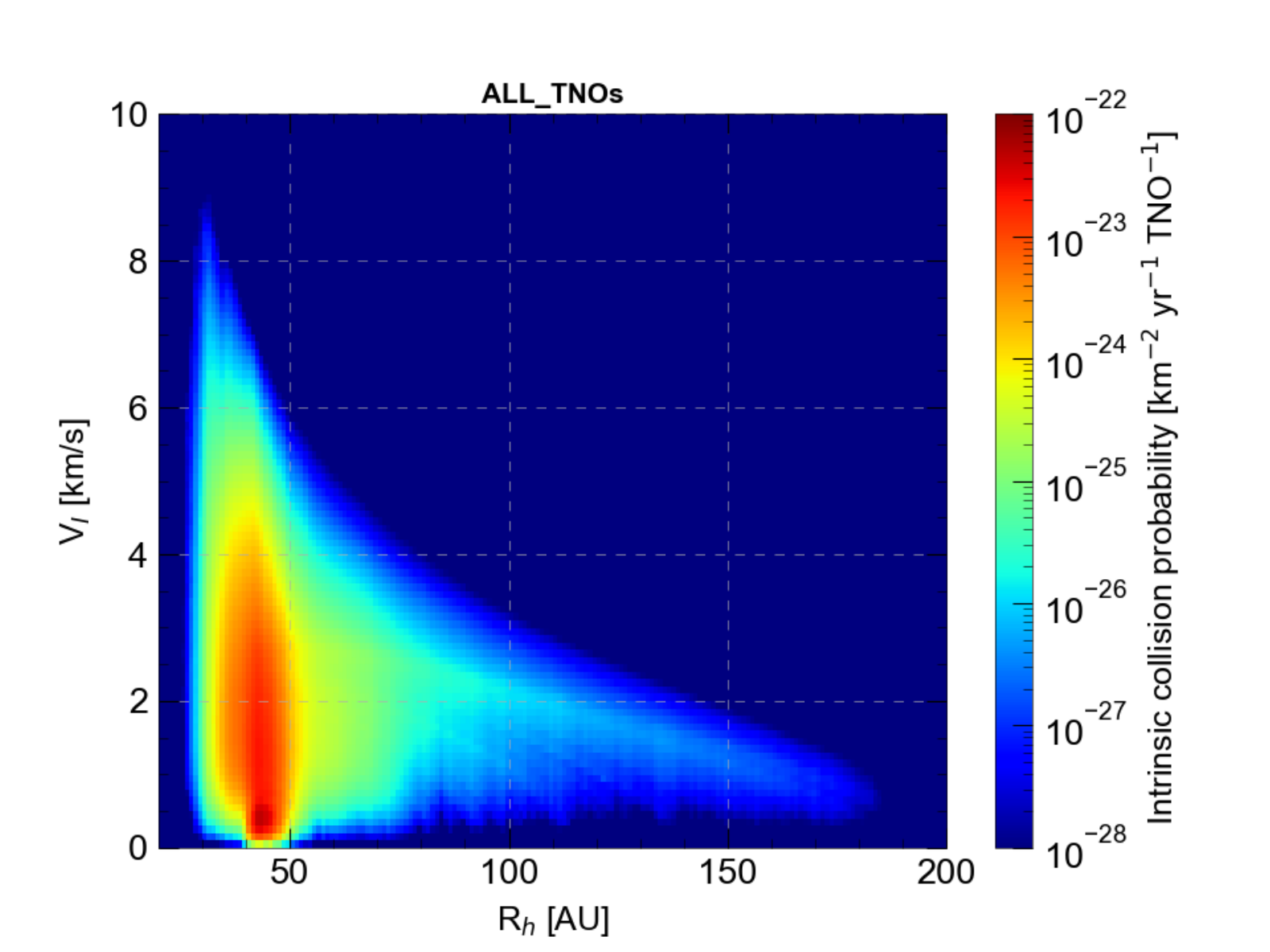}{0.45\textwidth}{(a)}
	\fig{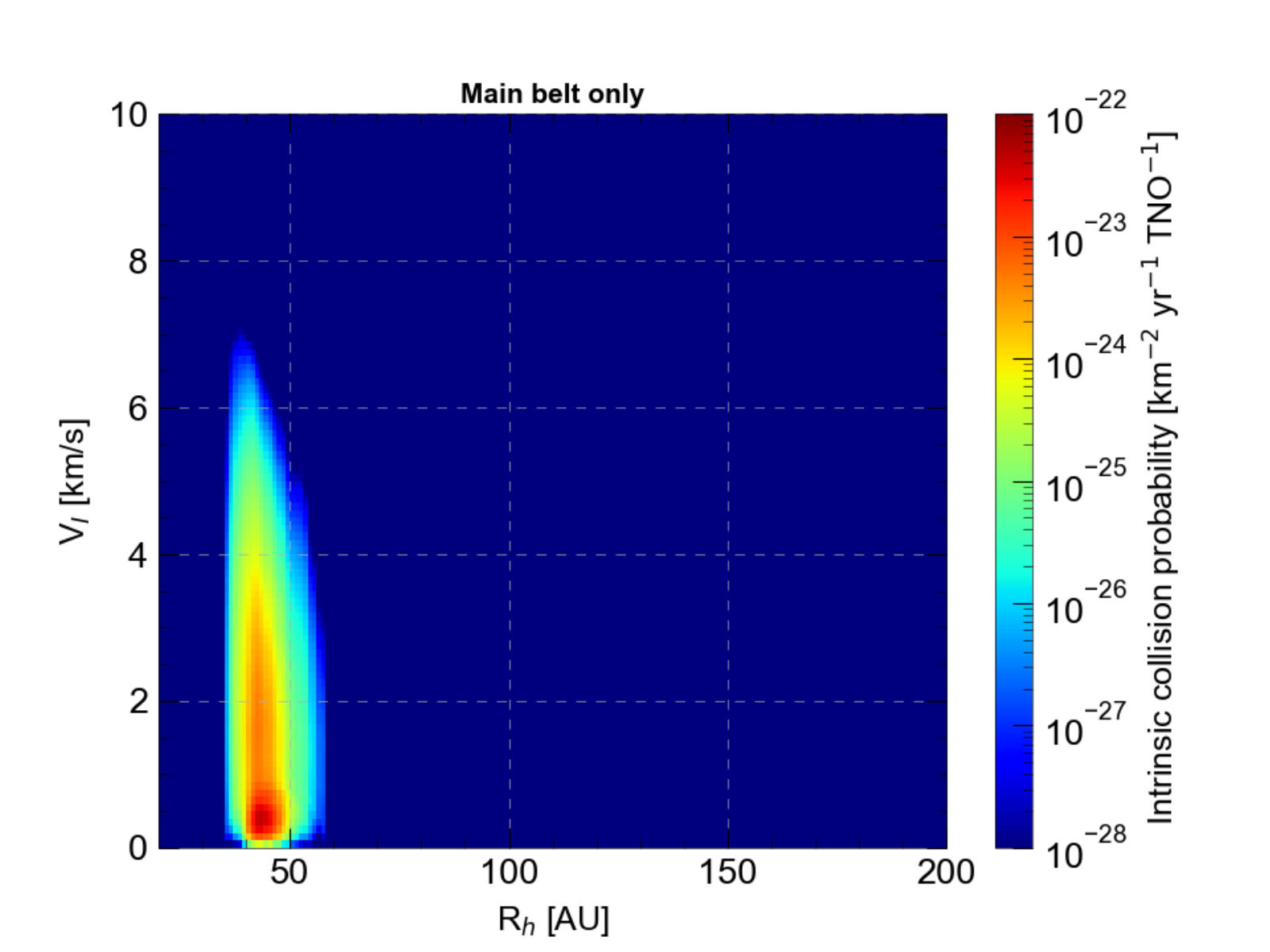}{0.45\textwidth}{(b)}}
	\caption{Combined collision probability (color bar) as a function of the heliocentric distance $R_h$ and impact speeds $V_I$, between all TNO populations (a) and the main Kuiper belt only (b). Subplot (a) is essentially the sum of all subplots in figures~\ref{fig:grid1}, \ref{fig:grid2} and \ref{fig:grid3}, while subplot (b) presents the intrinsic collision probabilities between "cold-cold", "hot-hot" and "hot-cold" subclasses. Each 2-D pixel, in each subplot, has dimensions of ($100$ m/s $\times$ 1~AU).}
	\label{fig:grid4}
\end{figure*}
\subsection{Combined Intrinsic Probabilities}

The $P_I$ for the Kuiper belt can now be determined by combining together the $P_I$ for the various sub-components determined in the previous section.  
To do this we must first weight each probability matrix by the relative size of the the components of the Kuiper belt population involved.  
Frustratingly, the actual relative sizes of these populations depends on the size range of sub-populations being considered as the `Excited` members of the Kuiper belt (the Hot Classical Kuiper belt, Resonant objects, Scattering disk and Outer/Detached components) appear to have one size distribution while the size-frequency distribution (SFD) of the Cold Classical Kuiper belt appears to be different.  
In Table~\ref{Tab:ncalcs} we summarize current literature population estimates of the various orbit classes in the Kuiper belt, down to the sizes of objects for which reasonably secure estimates of the populations are available.
We use the fractional population estimates of the sub-classes to combine together the $P_I$ grids for the individual components and determine a global map of $P_I$ for the Kuiper belt region (Figure~\ref{fig:grid4}).   

\begin{deluxetable}{lDrll}
  	\tablecaption{TNO sub-component population estimates.\label{Tab:ncalcs}}
\tablehead{
\colhead{Orbit Class} 
& \multicolumn2c{Fraction} 
& \multicolumn2c{Population} \\
& \multicolumn2c{\% of total} 
& \multicolumn2c{$N(H_r < 8.66)$}
}
\tablecolumns{3}
\startdata
\decimals
Cold Main Belt & 20 & 61000 & $\pm$6000  \\
Hot  Main Belt & 12 & 38000 & $\pm$9000  \\
Detached/Outer & 26 & 80000 & $\pm$40000  \\
Scattering  & 29 & 90000 & $\pm$20000  \\
\sidehead{Inner Resonances}
\tableline
 4:3 & 0.1 & 400 &$^{+800}_{-120}$  \\
 3:2 & 4.2 & 13000 &$^{+6000}_{-5000}$ \\
\sidehead{Main Resonances}
\tableline
5:3 & 1.6  & 5000 &$^{+5200}_{-3000}$ \\
7:4 & 0.9  & 3000 &$^{+4000}_{-2000}$ \\
\sidehead{Outer resonances}
\tableline
2:1 & 1.2  & 3700 &$^{+4400}_{-2400}$ \\
7:3 & 1.3  & 4000 &$^{+8000}_{-3000}$ \\
5:2 & 3.8  & 12000 &$^{+15000}_{-8000}$ \\
\enddata
\tablerefs{\citet{petit11},\citet{lawler18},\citet{gladman12}}
\end{deluxetable}

\section{Collision Rates\label{sec:results}}

The ``final" collision probability (yr$^{-1}$~TNO$^{-1}$) between two objects of radii $r$ and $R$ is obtained by multiplying Eq.~\ref{eq:intr} with the cross-sectional area of the two objects e.g. $P_F=P_I \times A = P_I(r+R)^2$. We remind the reader that the factor of ``$\pi$", in the cross-sectional area, is incorporated in $P_I$. 

To obtain the number of collisions $N_c$, per unit time, for an ensemble of objects in a given size range, $P_F$ needs to be integrated over a suitable size distributions, representing the number of objects relevant for these populations. For example, the number of collisions $N_c$, per unit time, that a single target of size $R=50$~km would experience, with a given KBO population in a given size range $r_1<r<r_2$, scales as the intrinsic collision probability between the target and objects from the given orbit class as well as the size and number of impactors per unit size in that class. 
In a given impactor size range, the number of collisions onto a single 50 km target will be: 
\begin{equation}\label{eq:ncoll2}
    N_c= P_I\int_{r_1}^{r_2}(50+r)^2\frac{dN}{dr}dr
\end{equation}
\noindent where $dN/dr$ is the number of impactors with size between $r$ and $r+dr$ for the given KBO orbit class. 
Where as if one is interested in the total number of collisions between all members of two orbit classes then one must integrate over both size distributions:
\begin{equation}\label{eq:sfd}
  N_c=P_I\int_r\int_R (r+R)^2 \frac{dN}{dr}\frac{dN}{dR}drdR\\
\end{equation}
where $dN/dr$ amd $dN/dR$ are the impactor and target size distributions.

If one is interested in the number of collisions over a given time interval, equation \ref{eq:sfd} is then multiplied by that time interval.

The SFD for Kuiper belt populations can be approximated using exponential distributions of the form $dN/dr \propto r^{-q}$. 
The size distribution does not follow a single power law  from the smallest to the largest bodies, but can be described using a small number of different logarithmic slopes, each valid for a given size range \citep[e.g.][]{bernstein04,petit08,petit11,fras14}. 
For a given impactor KBO population, the number of collisions $N_{ct}$, a target of $R=50$~km, would experiences over a given time $T$ will be the sum over the different logarithmic slopes appropriate to those impactors multiplied by the $P_I$ appropriate to the target and impactor orbital classes being considered. That is:
\begin{equation}\label{eq:ncoll-range}
    N_{ct}=T P_I\sum_i\left[\int_{{r_1}_i}^{{r_2}_i}(50+r)^2\left(\frac{r}{{r_{o}}_i}\right)^{-q_i}dr\right]
\end{equation}
where ${r_1}_i$ and ${r_2}_i$ is the range over which the logarithmic slope $q_i$ is valid and ${r_o}_i$ is the normalization. 
The total number of collisions can be obtained by adding the contribution from each KBO population:
\begin{align}
    \begin{split}
    N_t &= \sum_j {{N_c}_j} \\
       & =  T \sum_j {{P_I}_j}\sum_i\left[\int_{{r_1}_{ij}}^{{r_2}_{ij}}(50+r)^2\left(\frac{r}{{r_{o}}_{ij}}\right)^{-q_{ij}}dr\right]
\label{eq:ncoll-all}
    \end{split}
\end{align}
where ${N_{c}}_j$ denotes the contribution of a given KBO population to the number of collisions. Similarly, ${P_{I}}_j$ is the intrinsic collision probability between a given target and KBO population.  
 
To provide a sense of the frequency of collisions in the Kuiper belt and to compare our calculated collision probabilities to previous works \citep[e.g][]{green15,green16} we present an estimate of the total number of \textit{impacts} on to a $R=50$~km target.  Precise estimates of the total number of collisions a particular target of $R=50$~km would experience depends, naturally, on the orbital elements of the target and the impactor population it could collide with as well as the gravitational focusing and acceleration due to non-negligible mass of the target body.  By considering a small target ($R=50~\text{km}$) we are able to neglect gravitational focusing. For our orbit we use mean collision probabilities for two target populations. We compute the total number of impacts that a mean ``inner resonant object" (IRO, our combined 4:3 and 3:2 resonant populations) and a mean ``Cold Classical Kuiper Belt Object" (CCKBO) would experience in the current Kuiper belt.  The mean impact rates determined here should be roughly similar when compared to more direct computations for specific bodies.

In addition, considering a specific size range of impactors allows for approximate comparison with crater counts on Pluto-Charon system \citep{singer19} and Arrokoth \citep{spencer20}.
For Charon impactor size scales from crater size as $ d = 0.07D^{1.151}$ where $d$ and $D$ are the impactor and crater sizes in km \citep[S2]{singer19}.  
Using this relation we determine that the $1.4~\text{km} < D < 10~\text{km}$ craters measured on the smooth area of Charon's Vulcan Planum \citep[Table~S2]{singer19} correspond, roughly, to impactors of $100~\text{m}\lesssim d \lesssim 1~\text{km}$.
For Arrokoth the New Horizons team reported craters in the size range of $200~\text{m} < D < 1~\text{km}$, which translate to impactor sizes $2~\text{m}<r<10~\text{m}$. 
This choice of size range maps, roughly, to impact crater sizes ($1.4~\text{km} < D < 10~\text{km}$) \citep{singer19} observed on Charon and to ($0.2 < D < 1$~km) \citep{spencer20} on Arrokoth, using impact cratering scaling laws from \citep{zahnle03}, a relationship also adopted by \citep{green15,green16}.
To determine the total number of impactors in these size ranges requires knowing the SFD of objects in each impactor group. \citet{petit11} provides population estimates for $H_g<8$, which we convert to $H_r$ using the color index $(g-r)=1.0$ for the cold and $(g-r)=0.8$ for the hot TNO components \citep[e.g.][]{Schwamb19}, respectively. The impactor sizes considered here are within absolute magnitude range $17<H_r<32$, in the ``$r$"-band with the exact conversion dependent on albedo choice, discussed below. The number of objects between $8.66<H_r<17$ for both, high and low inclination TNOs was then estimated using a single logarithmic slopes of $\alpha=0.4$ or $(q=5\alpha+1=3)$, extrapolating smoothly from the $N(H_r<8.66)$ in Table~\ref{Tab:ncalcs}, i.e. we did not consider a \textit{divoted} size-frequency distribution \citep{shankman16} as was explored in \citet{green16,green19}.   
  
For the range $17<H_r<32$ we use an average slope of $\alpha=0.2 (q=2)$ based on the slope in the crater size-frequency distribution seen on Charon ($q=1.8\pm0.2$) \citep{singer19} and Arrokoth ($q=2.3\pm0.6$) \citep{spencer20}. To convert absolute magnitude to impactor and target size, albedos of 0.14 and 0.08 were assumed for the low and high inclination populations. 

\begin{figure*}[htpb]
	\gridline{\fig{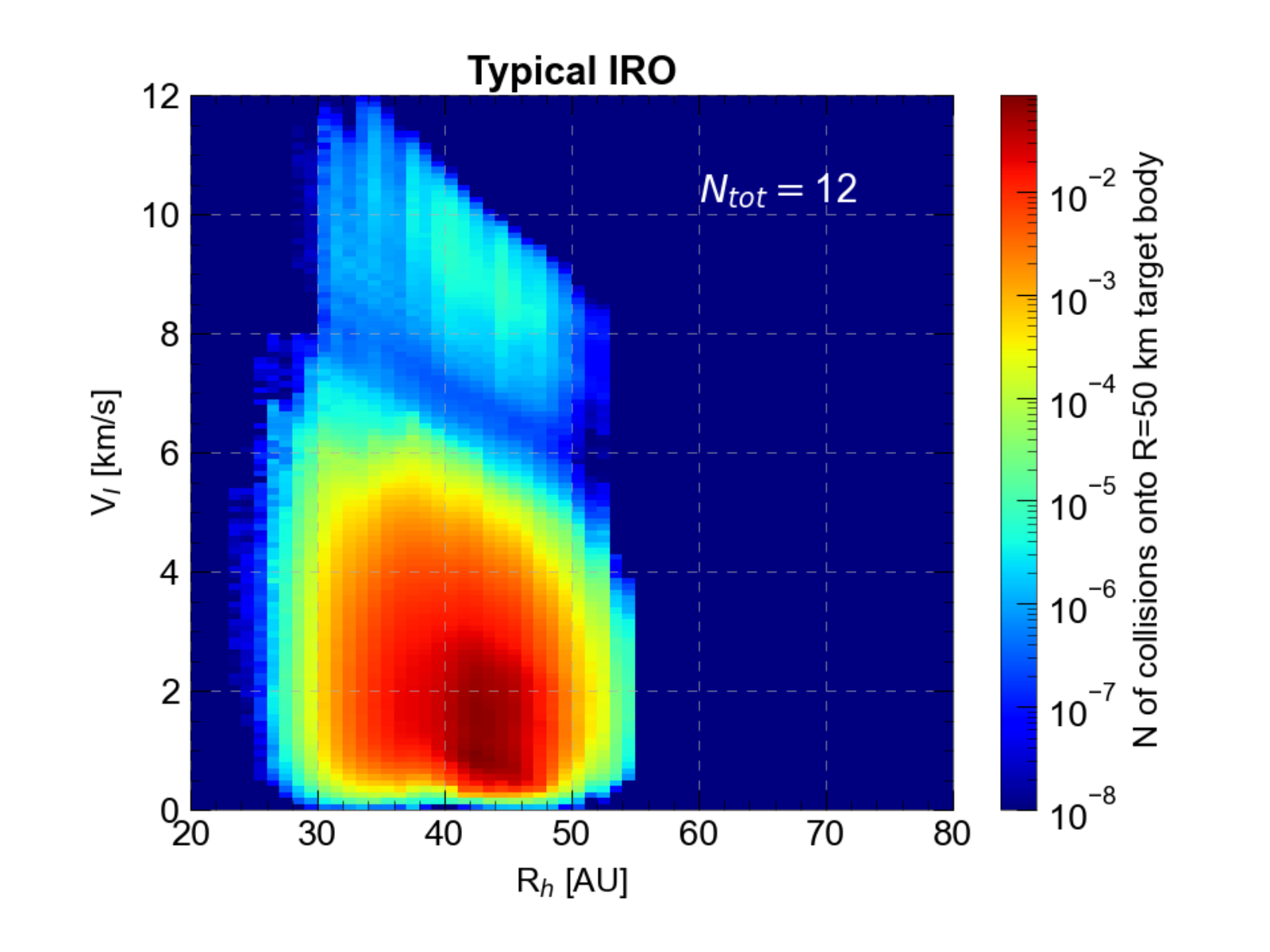}{0.45\textwidth}{(a)}
	\fig{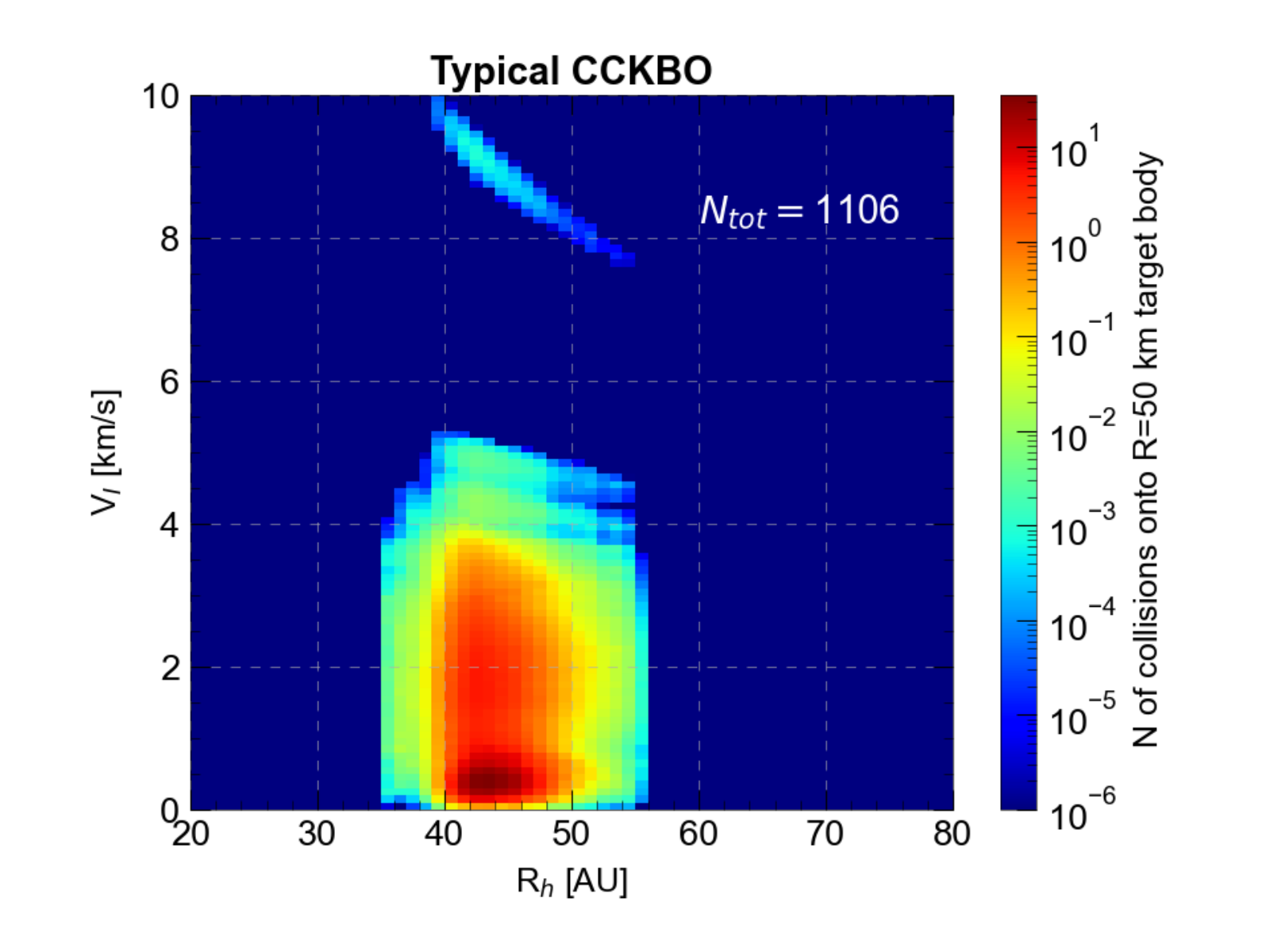}{0.45\textwidth}{(b)}}
	\caption{Number of collisions (color bar) onto an average ``plutino" (a) and onto a typical ``Cold Classical Kuiper Belt Object" (b). 
	In both cases the target is assumed to have radius $R=50$~km. The total number $N_{tot}$ of impacts are estimated over 4~Gyr, with impactors in the range $100~\text{m}<r<1~\text{km}$ (a) and $2~\text{m}<r<10~\text{m}$ (b). Each 2-D pixel, in each subplot, has dimensions of ($100$ m/s $\times$ 1~AU). The discontinuity and small contribution to high-speed impacts ($V_I>8$~km.s$^{-1}$ in both plots are attributed to the "granularity" in our scattering population model.)}
	\label{fig:grid5}
\end{figure*}

Using populations sizes and collisions probabilities for our sub-population groups we then compute the total impact rate onto our nominal IRO and CCKBO objects. Our model prediction of the number of collisions, distributed by velocity and heliocentric distance, that a single target of size $R=$50~km, belonging to the inner resonant and CCKBO populations would experience over the age of the Solar System are presented in Fig.~\ref{fig:grid5}a and Fig~\ref{fig:grid5}b. 
These impact distributions differ, somewhat, from each other and provide an indication of where within the Kuiper belt each sub-population is experiencing collisions and thus where those populations are most likely to be the source of local dust.  In particular we note that for the IRO orbit a significant number of impacts occur interior to 40~au and those impacts inthe 40~au to 45~au zone have speeds of between 0.7 km/s and 3 km/s while for the CCKBO almost no impacts occur interior to 40~au and the bulk of impacts are in the 40-50~au zone with impact speeds well below 1~km/s.

\begin{figure*}[htpb]
    \gridline{\fig{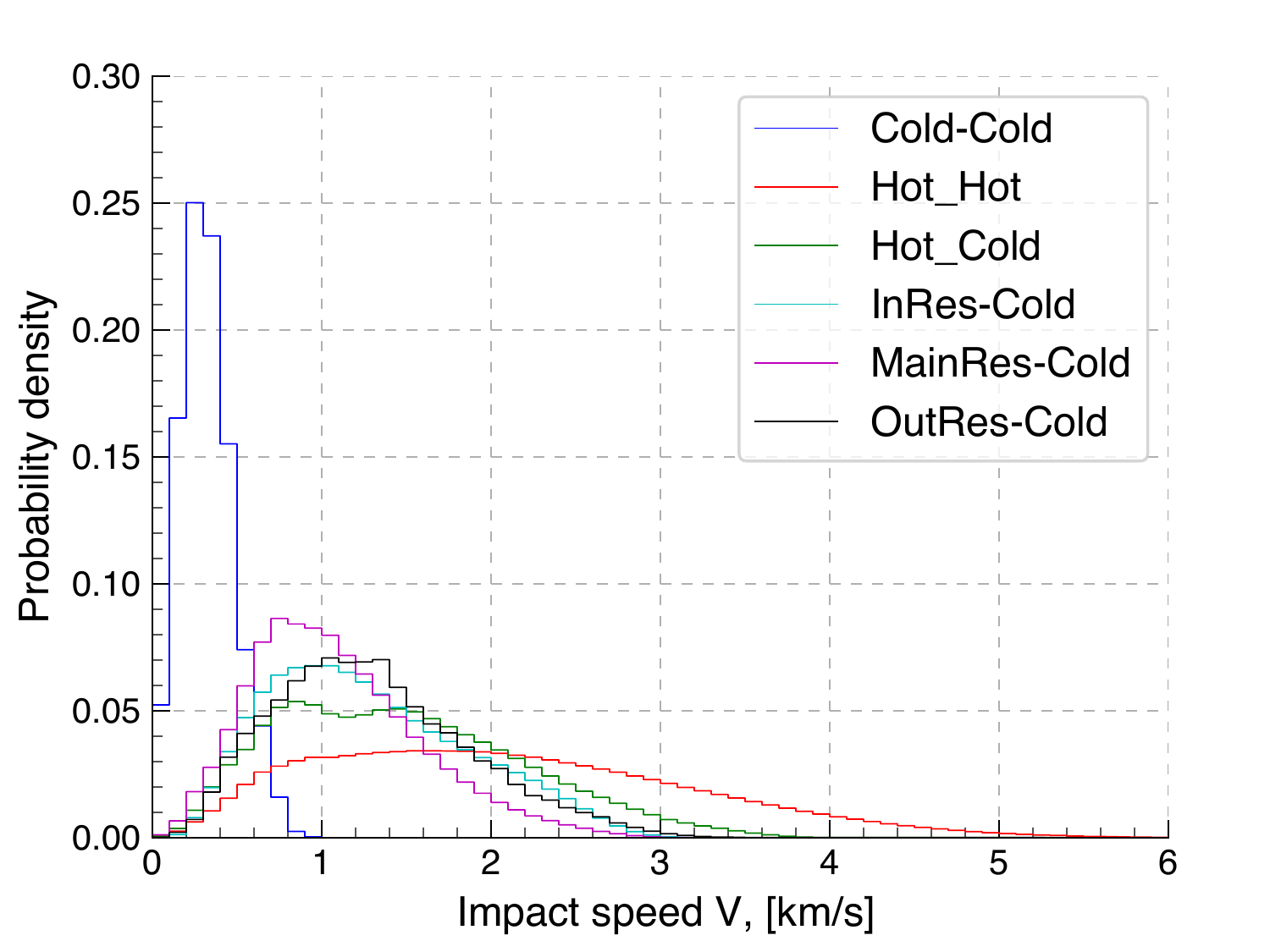}{0.45\linewidth}{a)}
       \fig{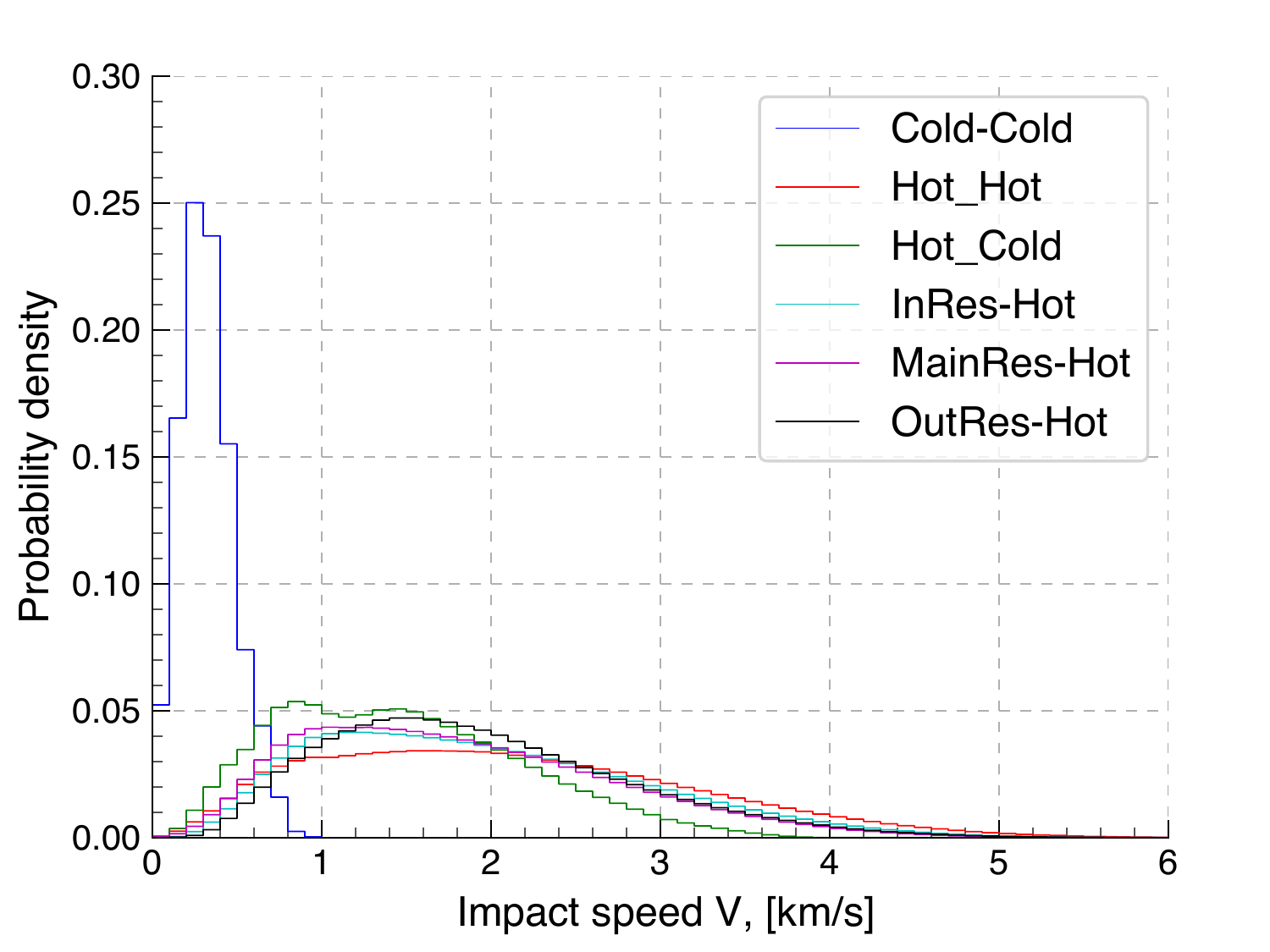}{0.45\linewidth}{b)}}
    \gridline{\fig{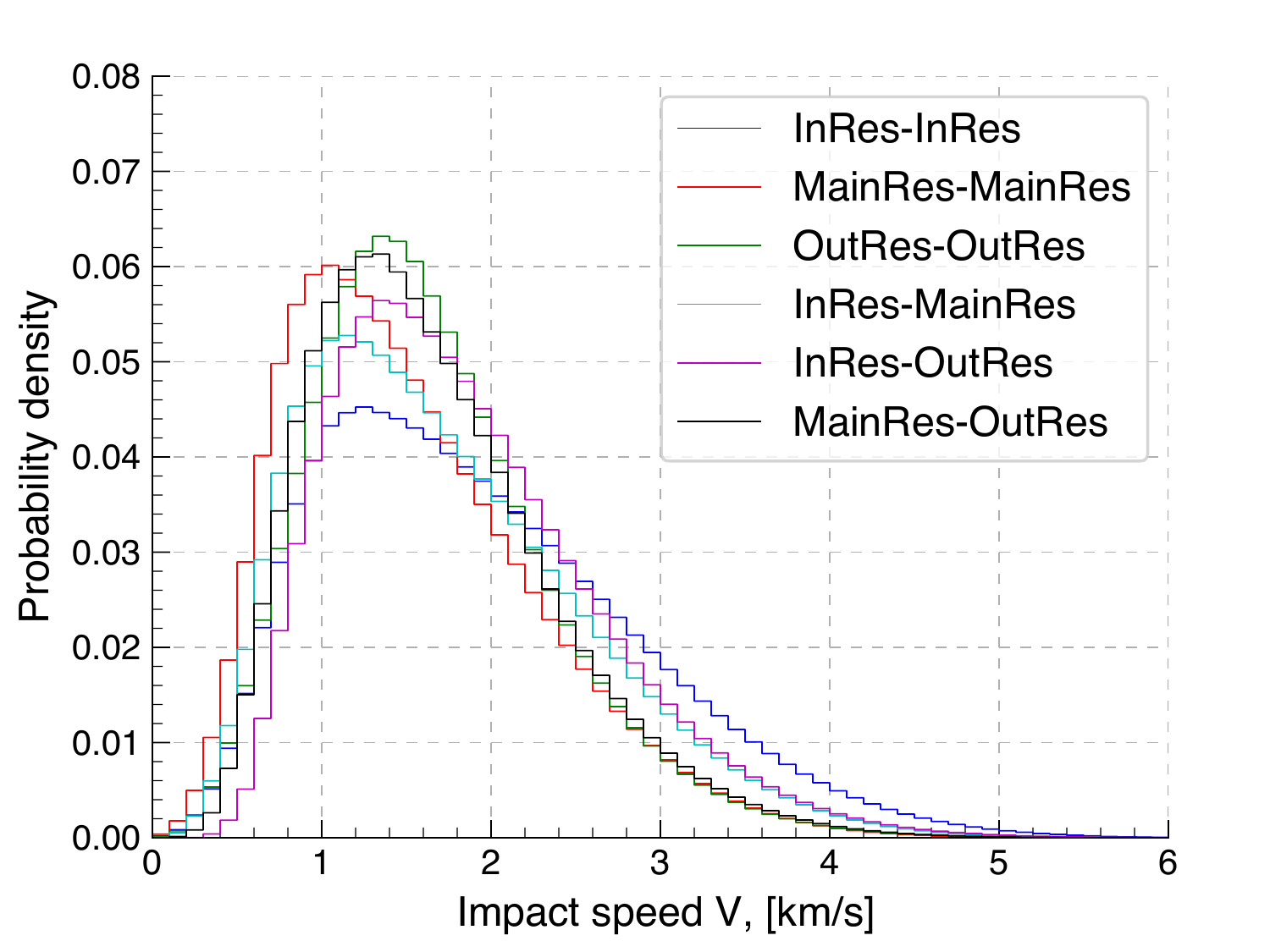}{0.45\linewidth}{c)}
       \fig{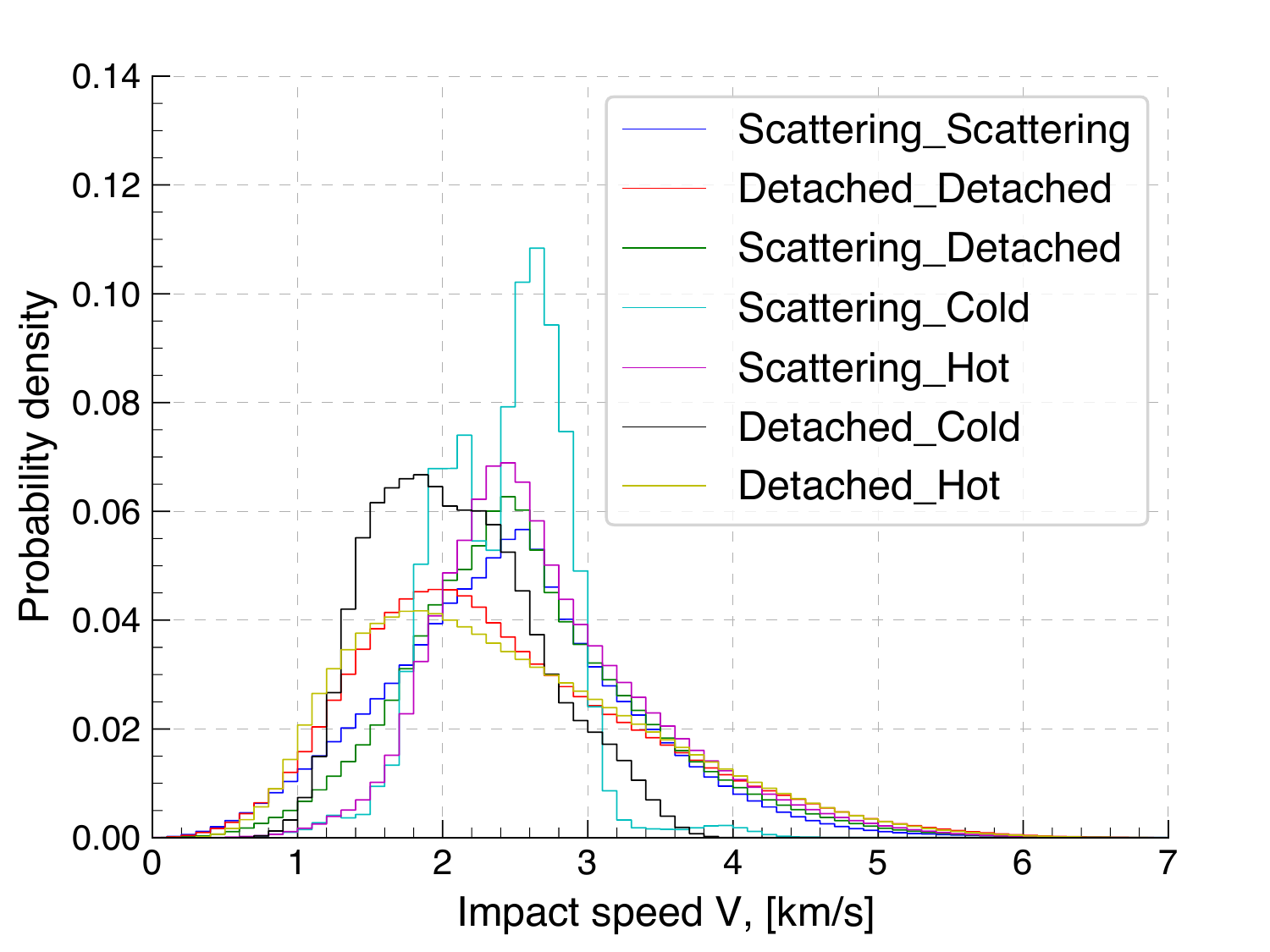}{0.45\linewidth}{d)}} 
    \gridline{\fig{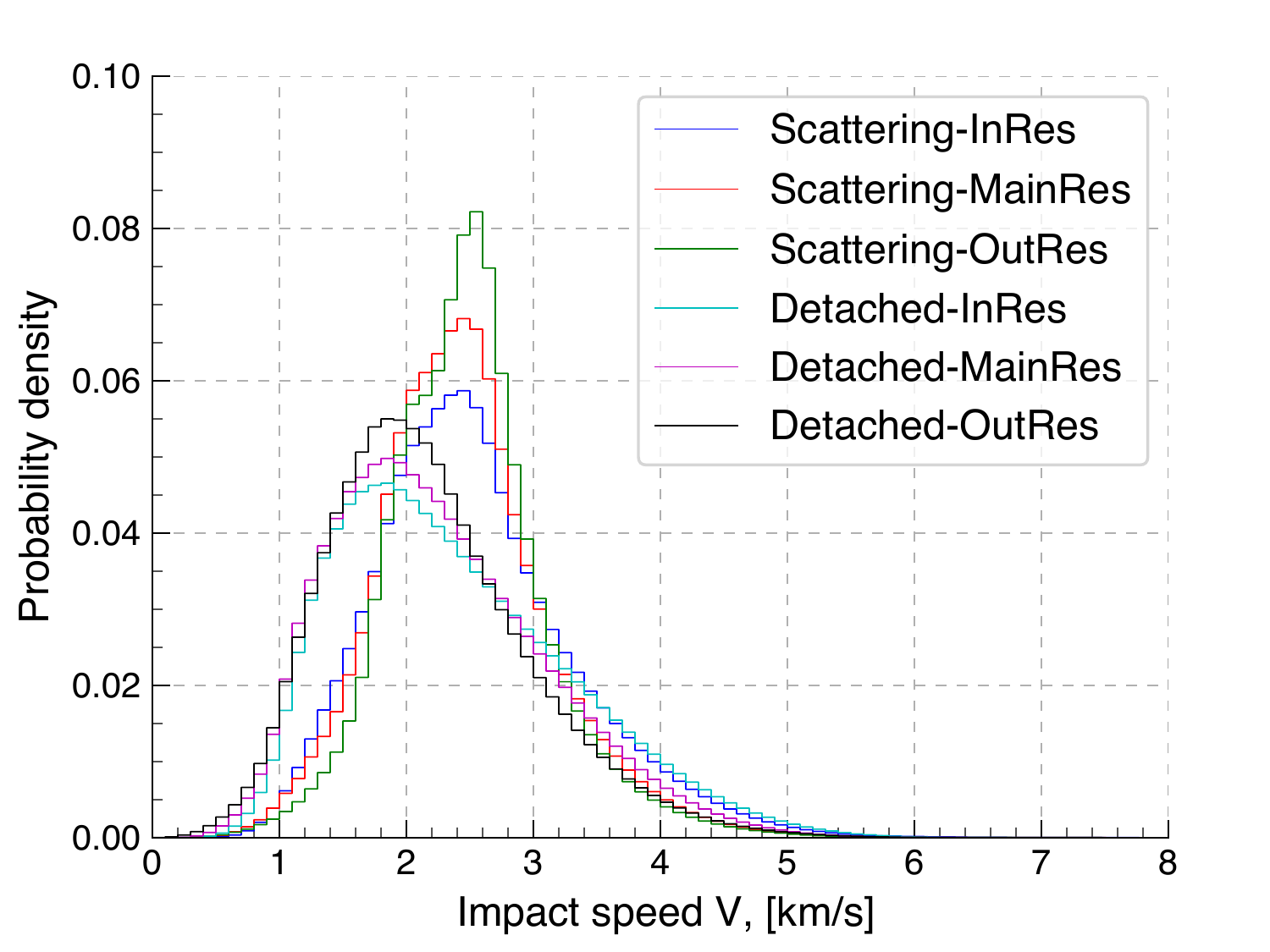}{0.45\linewidth}{e)}}    
	\caption{Probability density of the impact speeds between different dynamical TNO populations. All plots have been normalized to unity.}
	\label{figs:ispeeds}
\end{figure*}

For the case of a IRO, a 50~km spherical body (surface area $\sim$ 31,000~km$^2$), our model predicts $N_{tot}\sim$12 impacts over the age of the Solar System, with
impactors in the range 0.01$<r<$1~km (Fig.~\ref{fig:grid5}a). 
The New Horizons science team reported 70 craters with size $1.4<D<10$~km on Charon's \emph{Vulcan Planitia} smooth terrain (see supplementary material in \citep{singer19}). Vulcan Planitia roughly subtends an area of $A\sim 35,000$~km$^2$ \citep{singer19} or $\sim10\%$ larger than the surface area of 50~km target body. 
We find that the impact rate onto a typical IRO is dominated ($\sim 35\%$) by cold classical TNO population, followed by the hot ($\sim 21\%$) and detached ($\sim 17\%$) sub-groups. Similar trend was observed by the analysis of \citep{green15,green16}. We find that the inner resonance population accounts for only $\sim 14\%$ of the total impact rate on our nominal IRO.
The offset between our predicted number of impacts and observed number of craters on Charon, aside from population estimate and SFD uncertainties is further affected by ignoring the following points: 

\begin{enumerate}
    \item We note that our model is designed to calculate the collision probability within and between different groups of TNO populations, as opposed to using the exact orbit of Charon.
    \item In this work we ignore the full resonant and ``Kozai-Lidov" dynamics of Charon's orbit, although we do add a correction factor of 1.5 \citep{green15,green16} to the collision rates of plutinos onto Charon. Both of these effects will add to non-uniform precession of the orbital elements, as well limit the range the orbital elements can span over a precession cycle, effectively resulting in an increased rate of encounters with other TNOs.
    \item Failing to precisely model the relationship between impactor flux and number of observed craters. Impact scaling-laws suggest that, aside from the size of the colliding objects, there is a strong correlation between the impact speed, gravitational acceleration of the target and bulk densities of the target and impactor \citep[e.g.][]{hols02,hols03,zahnle03,hous11,hous18}. We convert observed impact crater diameters on Charon to impactor size, using an average collision speed of 2~km.s$^{-1}$ for all populations. However, fig.~\ref{figs:ispeeds} shows that the distribution of collision speeds, between some TNO populations, exhibit long tails with $V_I$ in excess of 3~km.s$^{-1}$. The effect of using an average impact speed, as opposed to a detailed distribution, in the crater to impactor size conversion will underestimate the size range of impactors considered in this work, which will translate to reduced total impactor flux. 
\end{enumerate}

For a detailed analysis of the impact rates onto a single object, the above shortcomings must be accounted for, as in \citep{green15,green16}. Recognizing the uncertainties, associated with TNO population estimates and in the SFD of objects smaller than 10~km, as well as neglecting the above mentioned points, our results cast confidence in the obtained collision probability calculations between different TNO classes. In a subsequent work, we will explore the actual collision rates in the trans-Neptunian region which urges for consideration of the above shortcomings, associated with our sample collision rate calculations.

For the case of the CCKBO, a target of radius $R\sim50$~km (surface area $\sim 31,000$~km$^2$) is expected to be impacted $\sim$ 1100 times, with impactors in the size range $2~\text{m}<r<10~\text{m}$, over the age of the Solar System. New Horizons team has counted 43 craters, over a surface area of 700~km$^2$, on the encounter hemisphere of Arrokoth, with crater size range ($0.2<D<1$)~km \citep{spencer20}. But the surface area, used for crater count on Arrokoth (700~km$^2$), is only $2.3\%$ of that of a spherical 50~km target body, so our modeling predicts $N_{tot}=1100 \times 0.023 \approx 25$ craters on Arrokoth's crater count surface. Unlike the IRO case, we find a better agreement with observations. The reason is that the low inclination component of the main classical Kuiper belt is more tightly confined near the ecliptic, effectively resulting in higher collision probability within the cold population and with other TNO groups. Furthermore, Arrokoth is expected to be primarily impacted by the cold classical TNOs, which display a tighter impact speed probability distribution, where the assumed average impact speed $V_i=0.3$~km~s$^{-1}$ is good assumption for the overall spread of impact speeds. This is further supported by our finding that 56\% of the impacts onto our nominal CCKBO are attributed to the cold classicals. We also find that the hot population is the next dominant impactor flux on our typical CCKBO, contributing $\sim$17\% of the total impact rate, followed by the detached $\sim$10\% and inner-resonant $\sim$8\% populations. The population contributing the least to the total impact flux is the scattering population, accounting for $\sim$2\%.
  
Finally, based on our modelling and analysis we find that most TNO collisions occur in the main classical belt, with the intrinsic collision probability for this sub-component accounting for $\sim 30\%$ (see Tab.~\ref{Tab:cols}) of the average total collision probability $\langle P_I\rangle \approx 4.5\times 10^{-21}$~km$^{-2}$~yr$^{-1}$~TNO$^{-1}$ within the TNO population. 
Within the main classical belt, collisions are dominated by ``Cold" on ``Cold" impacts ($\sim$17~\%) with median impact speeds $\langle V_I \rangle \sim0.3$~km~s$^{-1}$. 
Next most important source of impacts are the resonant TNOs (taking all resonances together) which account for $\sim$26~\% of the total collision probability. Resonant TNOs experience collisions at speeds in excess of 1~km~s$^{-1}$ and may be more significant sources of dust than the impacts within the ``Cold'' population.  
In contrast, we find that the collision probabilities, within the scattering population, are almost four orders of magnitude lower, compared to that in the main classical belt (Tab.~\ref{Tab:cols}. See also Fig~\ref{fig:grid1}a and Fig.~\ref{fig:grid3}i). These results provide a strong proxy as to where in the Kuiper belt the dust production is expected to dominate. Models of dust production and general cratering among TNOs will need to take into account the differences in collisional probabilities between the two major sources of impacts.

\section{Discussion and Conclusions}\label{sec:discus}

We have calculated the intrinsic collision probabilities and the corresponding impact speeds between different TNO populations, based on a revised dynamical model of the Kuiper belt, derived from the OSSOS ensemble sample, finding that low-velocity impacts at distances $30 <R_h<50$~au are most likely.

Our calculations indicate that collisions are concentrated in the main Kuiper belt region (30-55~AU) but impacts over a wide range of distances (30-150~AU), are not excluded. Although our simulations indicate collision speeds in excess of 8~km.s$^{-1}$, most of the collisions are likely to happen with speeds well below $V_I<5$~km/s, over the age of the Solar System. 

The highest intrinsic collision probability, $P_I\sim 7.5\times 10^{-22}$~km$^{-2}$~yr$^{-1}$~TNO$^{-1}$, is observed for impacts within the cold main classical population, with these collisions occurring at the lowest median speed $\langle V_I \rangle \sim 0.3$~km.s$^{-1}$, and accounting for $\sim 17\%$ of the average total collision probability $\langle P_I\rangle \approx 4.5\times 10^{-21}$~km$^{-2}$~yr$^{-1}$~TNO$^{-1}$ of all TNO populations. The dominance of impacts in the cold classical belt is not particularly surprising as these objects spend more time physically proximate to each other and close to the ecliptic, compared to the hot classical, scattering and resonant objects whose orbits carry them over large ranges of heliocentric distance, resulting in lower overall space density within these orbital classes.
The low impact velocities among objects with the highest impact probabilities results from the narrower semi-major axis, eccentricity and inclination distribution of main classical KBOs, compared to the detached and scattering populations.  
Moreover, the highest impact speeds between the cold objects do not exceed 1~km.s$^{-1}$, raising a fair question as to whether these collisions are catastrophic or even crater forming. The outcome of a collision is a strong proxy as to the amount of material (dust) liberated in these impact events. Perhaps, due to the low impact speeds, collisions within the cold classical population are most likely to lead to crater formation rather than to a complete disruption of colliding bodies. Furthermore, the low impact speeds may also favor slow merging of cold KBOs into a single binary object. 

The next highest collision probability ($P_I=3.2\times10^{-22}$ km$^{-2}$~yr$^{-1}$~TNO$^{-1}$) is observed between the hot and cold populations, with contribution of $\sim 7\%$ of the total average intrinsic collision probability, though with noticeably higher impact speeds. In a subsequent work, we will investigate if these higher impact speeds are destructive, accounting for specific impact energy involved.  Although somewhat lower in probability, collisions between hot and cold classical KBOs may very well dominate the dust production in the Kuiper belt, given their impact speeds.

We see a bi-modality in the velocity distribution for collisions between the scattering and cold populations, which could be a sampling ``artifact" caused by the low-fidelity of our input scattering model.  A few outliers with high inclination and eccentricity of the scattering model distribution over-contributing to the probability distribution. The gap in probabilities is not evident for all orbital populations and tends to be more prominent for impactors with high eccentricity and semi-major axis dispersion, which the relative velocity is a strong function of. A few such orbits whose semi-major axis overlap due to numerical resolution can lead to granularity in the computed impact velocity distribution. 

The lowest collision probability appears to happen within the scattering and detached populations with median collision speeds in excess of 2~km/s. Collision between these objects could result in complete destruction, depending on the size of colliding bodies. Given the low-intrinsic probability and total number of objects on these orbits, the likelihood of collisions, even after 4~Gyr of evolution in the TNO population, are now very low. Collisions among these populations may have produced significant levels of dust, early in the Solar Systems history when the number objects on such orbits would have been 10s to 100s of times higher. 

Similar to \citet{delloro13}, we find that $ P_I$ is indeed lower than previously reported values e.g., \citep{davis97}, though our computed collision speeds slightly differ from \citet{delloro13}. We find that the cold main classicals collide on average with speeds $\langle V_I \rangle \sim 0.3$ km/s, about 40\% lower than the ones obtained by \citet{delloro13}. Similarly, our impact speed between the hot component of the main belt range from 0.6 - 4.4~km/s, with a median value of $\langle V_I \rangle \sim 2.0$~km/s, whereas the value reported by \citet{delloro13} is $\sim20\%$ higher. The most likely reason for that is that we report the median collision speed, based on the \"{O}pik and Wetherill's formalism of collisions between asteroids, whereas the aforementioned authors provide results from best fit to Monte-Carlo simulations.

We also provide an approximate expected number of impacts onto a 50~km sample plutino and CCKBO and compared them to the observed crater densities on Charon and Arrokoth, by the New Horizons spacecraft. We find that, our results display a good agreement with crater densities reported by New Horizons, and also in agreement with the results first predicted by \citet{green15, green16, green19}.

\bibliography{colprob}{}
\bibliographystyle{aasjournal}

\end{document}